\documentclass[%
 reprint,
amsmath,amssymb, aps,
]{revtex4-2}
\usepackage{multirow}
\usepackage{xcolor}
\usepackage{natbib}
\usepackage{subcaption}%
\usepackage{graphicx}
\usepackage{url}
\usepackage{adjustbox}
\begin{document}
\title{Arbitrage equilibria in active matter systems}
\author{Venkat Venkatasubramanian}%
\email{venkat@columbia.edu}
\affiliation{Complex Resilient Intelligent Systems Laboratory, Department of Chemical Engineering, Columbia University, New York, NY 10027}

\author{Abhishek Sivaram}
\affiliation{Department of Chemical and Biochemical Engineering, Technical University of Denmark, 2800 Kongens Lyngby, Denmark}

\author{N. Sanjeevrajan} 
\affiliation{Department of Materials Engineering, Indian Institute of Technology-Madras, Chennai, India 600036}

\author{Arun Sankar} 
\affiliation{School of Electrical, Computer, and Energy Engineering, Arizona State University, Tempe, AZ}

\maketitle

\section*{Abstract}
The motility-induced phase separation (MIPS) phenomenon in active matter has been of great interest for the past decade or so. A central conceptual puzzle is that this behavior, which is generally characterized as a nonequilibrium phenomenon, can yet be explained using simple equilibrium models of thermodynamics. Here, we address this problem using a new theory, \textit{statistical teleodynamics}, which is a conceptual synthesis of game theory and statistical mechanics. In this framework, active agents compete in their pursuit of \textit{maximum effective utility}, and this self-organizing dynamics results in an \textit{arbitrage equilibrium} in which all agents have the same effective utility. We show that MIPS is an example of arbitrage equilibrium and that it is mathematically equivalent to other phase-separation phenomena in entirely different domains, such as sociology and economics.  As examples, we present the behavior of Janus particles in a potential trap and the effect of chemotaxis on MIPS. 

\section{Introduction}

Active matter describes systems composed of a large number of self-actualizing dynamical agents that consume and dissipate energy and exhibit interesting macroscopic behaviors \cite{toner2005hydrodynamics, narayan2007long, ramaswamy2010mechanics, marchetti2013hydrodynamics,cates2013active, brady2015thermoActivematter, cates2015motility}. Biological examples of such self-organizing systems include bacteria, ants, birds, mussels, etc. Nonliving active matter examples include self-propelled Janus particles, layers of vibrated granular rods, and so on. A central conceptual puzzle in our evolving understanding of active matter is why and when a collection of active agents that looks like an out-of-equilibrium system on the microscopic scale behaves macroscopically like a simple equilibrium system of passive matter
\cite{cates2013active, cates2015motility, gonnella2015motility, brady2015thermoActivematter, berkowitz2020active, o2020lamellar}.\\

Recently, a new framework called \textit{statistical teleodynamics} has been developed to predict the macroscopic emergent behavior of active matter systems that comprise biological, ecological, and socioeconomic agents~\cite{venkat2015howmuch, venkat2017book, venkat2022unified, venkat2022garuds, venkat2023mussels, venkat2024social}. Statistical teleodynamics is a natural generalization of statistical thermodynamics for goal-driven agents in active matter. It is a conceptual synthesis of the central concepts and techniques of population games theory with those of statistical mechanics towards a unified theory of emergent equilibrium phenomena and pattern formation in active matter. The name comes from the Greek word \textit{telos}, which means goal. Just as the dynamical behavior of molecules is driven by thermal agitation (hence \textit{thermo}dynamics), the dynamics of purposeful agents are driven by the pursuit of their goals and hence \textit{teleo}dynamics. \\

The fundamental quantity in statistical teleodynamics is the \textit{effective utility} of an agent, which measures the net benefit of an agent after subtracting all the costs the agent incurred in acquiring it. All agents in a population \textit{compete} to increase their effective utilities. Population game theory proves that this competitive \textit{pursuit of maximum utility}, under certain conditions, leads to an equilibrium, called the \textit{arbitrage} equilibrium, where the effective utilities of all agents are equal~\cite{sandholm2010population}. \\ 

There is an important philosophical difference between statistical teleodynamics and statistical thermodynamics. Statistical teleodynamics acknowledges the importance of recognizing the individual active agent and its behavioral properties \textit{explicitly} in developing a \textit{bottom-up} analytical framework of emergent phenomena. It also accounts overtly for the role of an agent's \textit{purpose} (e.g., survival and growth) and the naturally attendant concept of the \textit{pursuit of maximum utility}. The role of purpose is not acknowledged in statistical mechanics, as there is no way to express that concept in that framework. However, it is the central concept in game theory and hence fits in naturally in statistical teleodynamics. \\

Since our theory is a bottom-up emergentist framework, it emphasizes the \textit{agent} perspective in contrast to statistical mechanics, which stresses the \textit{system} view. For example, whenever equilibrium is addressed in statistical mechanics, it is usually formulated in terms of minimizing free energy, which is a top-down  \textit{system} perspective. However, in statistical teleodynamics, while the system view is also present via the maximization of the game-theoretic potential (as we discuss in the following sections), the \textit{equality of effective utilities} for all agents as the equilibrium criterion from the \textit{agent} perspective is conspicuously recognized and exploited. \\

In our previous work, using statistical teleodynamics, we have shown that emergent self-organized behaviors of active agents in different disciplines, such as bacterial chemotaxis~\cite{venkat2022unified}, ant crater formation~\cite{venkat2022unified}, flocking of birds~\cite{venkat2022garuds}, mussel bed patterning~\cite{venkat2023mussels}, social segregation~\cite{venkat2024social}, and income distributions~\cite{venkat2017book}, can be understood as the result of arbitrage equilibria in their respective contexts. The self-actualizing agents in these studies are examples of living agents found in biology, ecology, sociology, and economics. Therefore, they are self-driven by survival purpose and the pursuit of maximum utility.\\ 

Although the statistical teleodynamics framework was developed to model and predict the emergent behavior of a large population of \textit{living} goal-driven agents, we believe that it could also be helpful in understanding the emergent behavior of \textit{nonliving} self-actualizing agents, such as Janus particles. Although nonliving agents are not purposeful, their self-actualizing behavior appears seemingly purposeful, as if they are pursuing some goal in their persistent dynamics. \\

In this paper, by applying the statistical teleodynamics framework, we show that the nonliving active matter systems are in arbitrage equilibria. Under certain mathematical conditions that we discuss in the rest of the paper, this arbitrage equilibrium is equivalent to a statistical or thermodynamic equilibrium, thereby resolving the above-mentioned conceptual puzzle. \\

The remainder of the paper is organized as follows. First, we introduce the statistical teleodynamics framework. Then, we show how the self-organizing dynamics of ant-crater formation is similar to the self-actualizing behavior of Janus particles in a potential trap~\cite{brady2016JanusTrap}, and the emergent distribution for both weak and strong traps is indeed an equilibrium distribution, a Weibull distribution. We then present a model game-theoretic system that predicts the emergent macroscopic behavior of a population of agents that spontaneously segregate under certain conditions at arbitrage equilibrium. We show how this arbitrage equilibrium is equivalent to motility-induced phase separation (MIPS)~\cite{cates2013active, cates2015motility, brady2015thermoActivematter, brady2023MechnicalMIPS}. We further extend this discussion to develop a utility-based model of chemotaxis-driven MIPS~\cite{zhao2023chemotactic}. In all these cases, we show that the final configurations are indeed in arbitrage equilibrium, which is equivalent to thermodynamic or statistical equilibrium.
 
\section{Statistical Teleodynamics, Potential Games, and Arbitrage Equilibrium}

As noted, statistical teleodynamics is a synthesis of the central concepts and techniques of population games theory with those of statistical mechanics. The theory of population games is concerned with predicting the final outcome(s) of a large population of goal-driven agents. Given a large collection of strategically interacting rational agents, where each agent is trying to decide and execute the best possible course of actions that maximizes the agent's~{\em payoff} or {\em utility} in light of similar strategies executed by the other agents, can we predict which strategies would be executed and what outcomes are likely~\cite{easley2010networks, sandholm2010population}? In particular, one would like to know whether such a game would lead to an equilibrium situation.\\
    
For some population games, one can identify a single scalar-valued global function, called a {\em potential} ($\phi(\boldsymbol{x})$), that captures the necessary information about the utilities {(where $\boldsymbol{x}$ is the state vector of the system)}. The {\em gradient} of the potential is the \textit{payoff} or \textit{utility}. Such games are called {\em potential games}~\cite{rosenthal1973class,sandholm2010population, easley2010networks, monderer1996potential}. A potential game reaches strategic equilibrium, called \textit{Nash equilibrium}, when the potential $\phi(\boldsymbol{x})$ is maximized. Furthermore, this equilibrium is unique if  $\phi(\boldsymbol{x})$ is strictly concave (i.e., $\partial^2 \phi /\partial^2 x < 0$)~\cite{sandholm2010population}. \\

In potential games, the utility $h_i$ of an agent in state $i$ is the gradient of potential $\phi(\boldsymbol{x})$, i.e.,
\begin{equation}
{h}_i(\boldsymbol{x})\equiv {\partial \phi(\boldsymbol{x})}/{\partial x_i}
\label{eq:utility-defn}
\end{equation}

where $x_i=N_i/N$ and $\boldsymbol{x}$ is the population vector. {$N_i$ is the number of agents in state $i$, and $N$ is the total number of agents}. Therefore, we have 
\begin{eqnarray}
\phi(\boldsymbol{x})&=&\sum_{i=1}^n\int {h}_i(\boldsymbol{x}){d}x_i \label{eq:potential}
\end{eqnarray}

where $n$ is the total number of states. \\

To determine the maximum potential, one can use the method of Lagrange multipliers with $L$ as Lagrangian and $\lambda$ as the Lagrange multiplier for the constraint $\sum_{i=1}^nx_i=1$:
\begin{equation}
L=\phi+\lambda(1-\sum_{i=1}^nx_i)
\label{eq:lagrangian}
\end{equation}

All agents enjoy the same utility in equilibrium, i.e., $h_i = h^*$. It is an \textit{arbitrage equilibrium} \cite{kanbur2020occupational} in which agents are no longer incentivized to switch states, as all states provide the same utility $h^*$. In other words, equilibrium is reached when the opportunity for arbitrage, i.e., the ability to increase one's utility simply by switching to another option or state at no cost, disappears. Thus, the maximization of $\phi$ and $h_i = h^*$ are equivalent when the equilibrium is unique (i.e., $\phi(\boldsymbol{x})$ is strictly concave \cite{sandholm2010population}), and both specify the same outcome, namely, an arbitrage equilibrium. The former stipulates it from the top-down system perspective, whereas the latter is the bottom-up agent perspective. Thus, this formulation exhibits the duality property.\\

Just as mechanical equilibrium is reached when the forces balance each other equally, thermal equilibrium is reached when the temperatures are equal, and phase equilibrium is achieved when the chemical potentials are equal, our theory demonstrates that a system of active agents will reach an arbitrage equilibrium when their effective utilities are equal.  Whenever the effective utility is of a particular mathematical form (as explained in Sections VI and VII), this game-theoretic Nash equilibrium is equivalent to the statistical Boltzmann equilibrium. In fact, our theory reveals the critical insight that both living and non-living agents are driven by arbitrage opportunities towards equilibrium, except that their \textit{arbitrage currencies} are different. For nonliving matter, the currency is the chemical potential, whereas for living matter, the effective utility. \\

Although the chemical potential (and free energy) is appropriate for describing nonliving physicochemical systems, its usage for living agents, such as bacteria, ants, birds, and so on, seems a bit awkward. We believe that effective utility (and the game-theoretic potential) is a more natural choice for active agents driven by survival and growth goals found in biology, ecology, sociology, and economics. Thus, the utility-oriented perspective helps us to extend the concepts and techniques of statistical thermodynamics more naturally to teleological agents by smoothly connecting with game theory. This is what has been accomplished by statistical teleodynamics.\\

We wish to emphasize that we are not claiming that the lower forms of living agents such as bacteria and ants pursue the survival goal and strategies \textit{rationally}. Our view is that the biological survival instincts of such agents cause particular dynamical behaviors that evolved over millions of years to help them improve their survival chances. Therefore, they act in a goal-driven manner \emph{instinctively}, which can be modeled using our framework of the pursuit of maximum utility or survival fitness.\\

Our goal is to identify the fundamental principles and mechanisms of self-organization of goal-driven agents. Towards that, we develop simple models that offer an appropriate coarse-grained description. The spirit of our modeling is similar to that of the van der Waals or the Ising model in statistical thermodynamics.  

\section{Ant crater model and Janus particles}

As an example of an active matter system, the dynamical behavior of self-actualized Janus particles in a potential trap has attracted considerable attention~\cite{brady2016JanusTrap, brady2016MipsCurrent}. Takatori et al.~\cite{brady2016JanusTrap} discuss the behavior of Janus particles in two regimes: (i) weak trap ($\alpha < 1$) and (ii) strong trap ($\alpha > 1$). They observe that the particle behavior in the weak regime can be considered an equilibrium outcome, whereas, in the strong regime, they suggest a nonequilibrium behavior. \\

We approach this system from the perspective of statistical teleodynamics. For us, this system resembles the behavior of a large population of ants building an ant colony underground. The dynamics of this activity involves the transport of sand grains by ants from an underground nest to the surface. The resulting grain pile, called the ant crater, is of a particular shape known as the Weibull distribution~\cite{venkat2022unified}. \\

Since grain transport involves an effort that increases with the distance the ant travels, the ants would prefer to drop the grains sooner rather than later to minimize the effort. However, if they drop them too close to the nest, the sand grains pile could collapse back into the nest, which would mean more work for them later on. Therefore, ants innately balance the need to transport grains as far away as possible while trying to minimize the effort (i.e., the disutility) expended in doing so.\\ 

As we showed recently~\cite{venkat2022unified}, this process can be modeled considering the effective utility of ants and their self-organizing competitive dynamics. In our model, the utility of an ant is determined by three factors. The first factor is the utility or benefit that it gains from having a home, the nest, given by $b > 0$. \\

The second factor describes the disutility (i.e., the cost) it incurs by transporting the grains away from the nest. We assume that the ants move outward radially from the nest with some average velocity $v$. We model the rate at which the ants drop off the grains as $s r^{a-1}$, where $r$ is the distance it travels from the nest to the drop-off point ($s > 0$ and $a > 1$ are constant parameters). The disutility of the effort $W$ an ant expends then depends on how much grain it carries and for how long. This results in
\begin{eqnarray*}
    W = \int_0^t sr^{a-1} d t = \int_0^r sr^{a-1} \frac{d r}{v} = \frac{sr^a}{va} = \frac{\omega r^a}{a}
\end{eqnarray*}
where $\omega = s/v$. In chemical engineering, $W$ is known as the Damk\"ohler number which quantifies the ratio of the time scales of flow to that of the reaction. Higher values of Damk\"ohler number, in this case, suggest a higher propensity of an ant to drop off the grains. \\
 
The third factor accounts for the disutility of competition among ants. As ants ($N_i$) try to crowd at the same location ($r_i$) to drop off their grains, this term forces them to spread out to minimize the cost of the competition. As Venkatasubramanian et al.~\cite{venkat2022unified} discuss, this term is modeled as $-\ln N_i$. \\

Combining all three, the effective utility $h_i$ that an ant gains by dropping a grain at a distance $r_i$ is given by
\begin{equation}
    h_i(r_i, N_i) = b - \frac{\omega r_i^a}{a} - \ln N_i 
    \label{eq:ant-utility}
\end{equation}

The potential for this system then becomes
\begin{eqnarray}
    \phi(\mathbf{x}) &=& \sum_{i=1}^n \int h_i(\mathbf{x}) d x_i\\
    &=& b - \frac{\omega}{a} \langle r^a \rangle + \frac{1}{N} \ln \frac{N!}{\prod_{i=1}^N (Nx_i)!} 
    \label{eq:ant-potential}
\end{eqnarray}
where $\langle r^a\rangle$ is the expectation of the quantity $r^a$, based on the locations of the ants ($N$ is the total number of ants). As the reader might recognize, the last term in Eq. 6 is entropy. Therefore, by maximizing potential $\phi$, one is equivalently maximizing entropy subject to the constraints in the first two terms. This deep connection between statistical mechanics (through entropy) and game theory (through potential) has been discussed in great detail here~\cite{venkat2017book, kanbur2020occupational}. We discuss this connection at some length in Section V. \\

As Venkatasubramanian et al.~\cite{venkat2022unified} show, there is a unique arbitrage equilibrium outcome for this collective behavior, where all ants have the same utility, i.e., $h_i = h^*$. Therefore, we have 
\begin{equation}
    h^* = b - \frac{\omega r_i^a}{a} - \ln N_i^* 
    \label{eq:ant-utility-equil}
\end{equation}

which can be rearranged to show
\begin{eqnarray}
x_i^* = \frac{N_i^*}{N} = \frac{\exp\left(-\dfrac{\omega r_i^a}{a} \right)}{\sum_j \exp\left(-\dfrac{\omega r_j^a}{a}\right)} 
\label{eq:ant-x_i}
\end{eqnarray}

where $N_i^*$ is the value in equilibrium. In the continuum limit, the states are continuous, where the state is defined as the radial location $r$. This results in the simplification $x_i^* = N_i^*/N =  \rho^*(r)2\pi r d r/N$, where $\rho^*(r)$ is the number density of ants at location $r$ at equilibrium. With this result in Eq~\eqref{eq:ant-x_i}, it can be shown that the number density of ants follows the distribution, 
\begin{equation}
    \rho^*(r) = \frac{A}{r} \exp\left({-\frac{\omega r^a}{a} }\right)
    \label{eq:ants_2}
\end{equation}

where $A$ is a constant that satisfies the boundary condition of a constant flux of ants from the center of the nest. \\

Note that this emergent distribution is that of the number of ants. Given this distribution of $\rho^*(r)$, the grain distribution can be calculated using a cumulative distribution given by
\begin{eqnarray*}
F(r) = \frac{\int_0^r (s r^{a-1} \rho^*) 2\pi r ~d r}{\int_0^{{\infty}} (s r^{a-1} \rho^*) 2\pi r ~d r}
\end{eqnarray*}

This gives,
\begin{eqnarray*}
    F(r) = 1 - \exp{\left(- \frac{\omega r^{a}}{a} \right)}
\end{eqnarray*}
with the grain distribution $f(r)$, given by
\begin{eqnarray}
    f(r) = \frac{d F}{d r} = \omega r^{a-1} \exp\left(-\frac{\omega r^{a}}{a}\right)
\end{eqnarray}

which is the Weibull distribution of ant craters that is observed empirically.  \\

It is important to emphasize that this distribution results from an \textit{equilibrium}, namely, the \textit{arbitrage equilibrium}. This is not a nonequilibrium or far-from-equilibrium outcome. 

\subsection{Janus particles in a potential trap}

We now consider the Janus particles from this perspective. We will show that the Janus particle dynamics can be considered as an arbitrage equilibrium outcome in both the weak and strong traps. \\

Our analysis is based on the work of Takatori et al.~\cite{brady2016JanusTrap}. They report on the behavior of self-propelled Janus particles that are in an acoustic trap whose strength can be tuned. The trap force is modeled by 
\begin{equation}
    F^{trap}(r) = -kr \exp (-2(r/w)^2)
    \label{eq:trap-force}
\end{equation}

where $k$ is the spring constant and $w$ is the width of the trap. They show that the probability distribution $P(r)$ due to the active Brownian motion of the swimmer is given by 
\begin{equation}
    P(\bar r)(U_0\tau_R)^2 = (\alpha/\pi) \exp(-\alpha \bar r^2)
    \label{eq:trap-p(r)}
\end{equation}

where $\bar r = r/(U_0\tau_R)$ and $\alpha := k\tau_R/\zeta$ is the nondimensional trap.\\

In our approach, we consider the ``struggle" of Janus particles swimming against the trap force to be similar to the effort expended by ants transporting sand grains. Therefore, we propose that the effective utility $H_i$ for a Janus particle be
\begin{equation}
    H_i(r_i, N_i) = - \frac{\omega r_i^a}{a} - \ln N_i 
    \label{eq:janus-utility}
\end{equation}

where the first term is the disutility of the work done by the Janus particle against the trap force, and the second term is the disutility of the competition among the particles, as before. We do not need the benefit term $b$ here, as it is not relevant for Janus particles. Following the same analysis for the dynamics of the ants, we conclude that the Janus particles will also reach the same arbitrage equilibrium outcome of $H_i = H^*$ with the probability distribution given by the Weibull distribution
\begin{eqnarray}
    P(r) = \omega r^{a-1} \exp\left(-\frac{\omega r^{a}}{a}\right)
    \label{eq:janus-weibull}
\end{eqnarray}

Now, we expect that the exponent $a$ would depend on the trap strength $\alpha$. As the trap strength increases, the work done by the particles against the trap force increases super linearly with distance $r$. We postulate that this nonlinear dependence can be modeled as 
\begin{eqnarray}
    a = 1 + \alpha + 0.5\alpha^2
    \label{eq:janus-alpha}
\end{eqnarray}

\begin{figure}[!ht]
\includegraphics[width=\linewidth]
{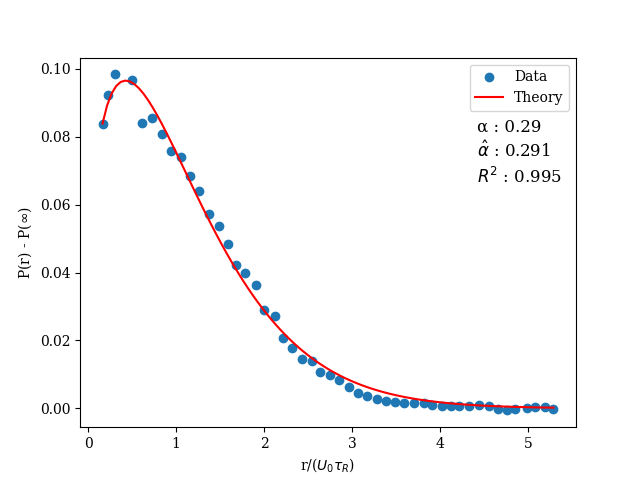}
\caption{{Weibull distribution fit for the weak trap: Experimental data}}
\label{fig-Weibull-weak-expt}
\end{figure}

\begin{figure}[!ht]
\includegraphics[width=\linewidth]
{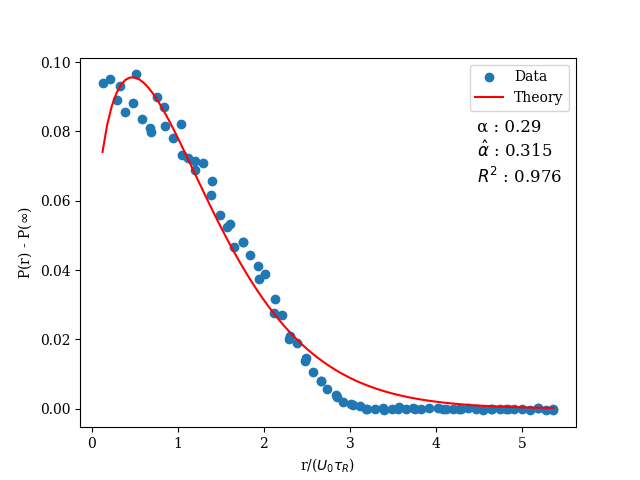}
\caption{{Weibull distribution fit for the weak trap: Simulation data}}
\label{fig-Weibull-weak-simu}
\end{figure}

\begin{figure}[!ht]
\includegraphics[width=\linewidth]
{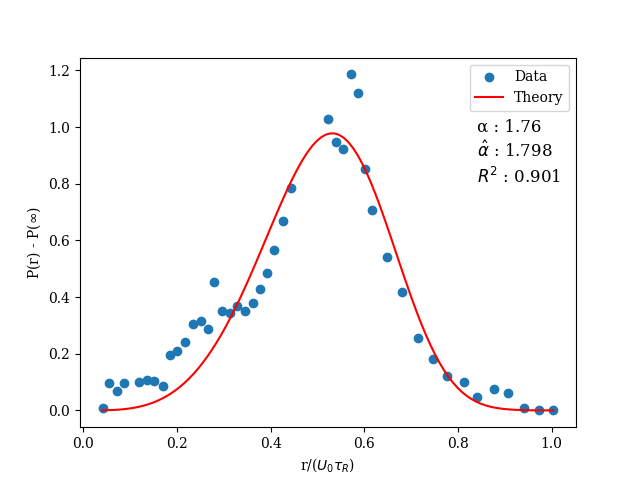}
\caption{{Weibull distribution fit for the strong trap: Experimental data}}
\label{fig-Weibull-strong-expt}
\end{figure}

\begin{figure}[!ht]
\includegraphics[width=\linewidth]
{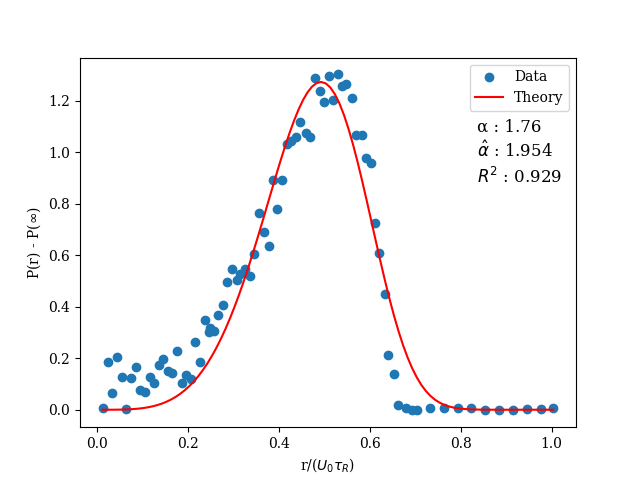}
\caption{{Weibull distribution fit for the strong trap: Simulation data}}
\label{fig-Weibull-strong-simu}
\end{figure}

We obtained the best fit for Eq.~\ref{eq:janus-weibull} for the weak ($\alpha <1$) and strong ($\alpha >1$) trap force data reported by Takatori et al.~\cite{brady2016JanusTrap} (Figs. 2a and 2b in their paper). The best-fit plots are shown in Figs.~\ref{fig-Weibull-weak-expt}-\ref{fig-Weibull-strong-simu}. As we can see, the Weibull distribution fits both regimes (weak and strong) quite well (the $R^2$ values are reported in the figures). We determined the exponent $a$ from the fitted distributions and used Eq.~\ref{eq:janus-alpha} to estimate the $\hat \alpha$ values. We see that the estimated values ($\hat{\alpha}$ shown in the figures) are in good agreement with the actual $\alpha$ values reported by Takatori et al.~\cite{brady2016JanusTrap}. \\

Thus, the data appear to support our prediction that the final distributions of the Janus particles in both the weak- and strong-trap regimes are the results of arbitrage equilibria. They are not out-of-equilibrium or nonequilibrium systems, but arbitrage equilibrium systems. More experimental and simulation studies are needed at other values of $\alpha$ to confirm this more conclusively. In addition, it would be helpful to derive the first term on the right-hand side of Eq.~\ref{eq:janus-utility}, and Eq.~\ref{eq:janus-alpha}, from the first principles. 

\section{Schelling-agents model and MIPS}

We now consider another game-theoretic model system to address motility-induced phase separation (MIPS) from a statistical teleodynamics perspective~\cite{venkat2022unified}. Here again, we demonstrate the emergence of arbitrage equilibrium as a result of the self-organizing dynamics of a large population of active agents competing for benefits. There are numerous examples of such behavior in the living world. In ecology, for example, we have the formation of mussel beds in the sea~\cite{koppel2005scale, liu2013CahnHilliard,deJager2020patterning,venkat2023mussels}; in sociology, the social segregation of different groups~\cite{schelling1969segregation,nilforoshan2023human}; and in economics, the emergence of income inequality in societies~\cite{venkat2017book}. \\

We develop our model by considering a large lattice of local neighborhoods or blocks, each with $M$ sites that agents can occupy. There are $n$ such blocks, $nM$ sites, and a total of $N$ agents, with an average agent density of $\rho_0 = N/(nM)$. The state of an agent is defined by specifying the block $i$ in which it is located, and the state of the system is defined by specifying the number of agents, $N_i$, in block $i$, for all blocks ($i \in \{1, \dots, n\}$ ). Let block $i$ also have $V_i$ vacant sites, so $V_i = M - N_i$.  \\

Next, as we did in the case of the ant crater-Janus particles, we define the effective utility, $h_i$, for agents in block $i$, which agents try to maximize by moving to better locations (i.e., other blocks), if possible. The effective utility is, again, the net sum of the benefits minus the costs, but the benefits and costs here are different from those of the ant-crater situation, as one might expect. \\

Here, an agent prefers to have more members in its neighborhood, as this aggregation provides certain benefits. In ecology, for example, for mussels, it improves the chances of survival of mussels against predators and sea wave stress~\cite{deJager2020patterning, liu2013CahnHilliard}. In sociology, it increases social benefits such as meeting life partners, finding better jobs, etc.~\cite{nilforoshan2023human}. Therefore, this \textit{affinity benefit} term, which represents cooperation between agents, is proportional to the number of agents in its neighborhood. We model this as $\alpha N_i$, where $\alpha >0$ is a parameter. \\

However, this affinity benefit comes with a cost. As more and more agents aggregate, they all compete for the same limited resources in the neighborhood. This \textit{congestion cost} is widely modeled by the quadratic disutility term $-\beta N_i^2$ ($\beta >0$)~\cite{venkat2017book, chone2019quad_utility}. \\

In addition, agents also balance two competing search strategies - \textit{exploitation} and \textit{exploration}. Exploitation takes advantage of the opportunities in the immediate, local neighborhood of the agents. On the other hand, exploration examines possibilities outside. This is a widely used strategy in biology, ecology, and sociology. For example, a genetic mutation can be thought of as exploitation, which is searching locally in the design space, while crossover is exploration, which is searching more globally.\\


Regarding exploration, the agents derive a benefit from having a large number of vacant sites to potentially move to in the future should such a need arise. This is the instinct to explore other opportunities, as new vacant sites are potentially new sources of food and other benefits. We call this resource the \textit{option benefit} term as agents have the option to move elsewhere if needed. Again, following Venkatasubramanian~\cite{venkat2017book, kanbur2020occupational, venkat2022unified}, we model this as $\gamma \ln (M-N_i)$,  $\gamma >0$. The logarithmic function captures the diminishing utility of this option, a commonly used feature in economics and game theory. As before, this benefit is also associated with a cost due to competition among agents for these vacant sites. As in the ant-crater case, we model this \textit{competition cost} as $-\delta \ln N_i$, $\delta >0$~\cite{venkat2015howmuch, venkat2017book, kanbur2020occupational}.\\

Combining all these, we have the following effective utility function $h_i$ for the agents in block $i$ as, 
\begin{eqnarray}
    h_i(N_i) = \alpha N_i  - \beta N_i^2 + \gamma \ln(M - N_i) - \delta \ln N_i
\end{eqnarray}

Intuitively, the first two terms in the equation model the benefit-cost trade-off in the exploitation behavior while the last two model a similar trade-off in exploration. \\

Rewriting this in terms of the density ($\rho_i$) of agents in block $i$, $\rho_i = N_i/M$, and absorbing the constant $M$ into $\alpha$ and $\beta$, we have
\begin{eqnarray}
    h_i(\boldsymbol \rho) = \alpha \rho_i - \beta \rho_i^2 + \gamma\ln (1- \rho_i) - \delta\ln \rho_i
    \label{eq:schelling-utility-non-simplified}
\end{eqnarray}

Note that in certain cases, agents may not occupy all $M$ sites and only a fraction of sites can be occupied due to restrictions such as steric factors. This corresponds to a maximum occupancy density ($\rho_{\text{max}}$). In such a situation the utility of vacant sites will be $\ln(1- \rho_i/\rho_{\text{max}})$, resulting in the formulation, 
\begin{eqnarray}
    h_i(\boldsymbol \rho) = \alpha \rho_i - \beta \rho_i^2 + \gamma\ln (1- \rho_i/\rho_{\mathrm{max}}) - \delta\ln \rho_i
    \label{eq:schelling-utility-non-simplified}
\end{eqnarray}

We can set $\delta = 1$ without any loss of generality. In addition, we set $\gamma = 1$ and $\rho_{\mathrm{max}}=1$ to gain analytical simplicity, but these can be relaxed later if necessary. Therefore, we now have 

\begin{eqnarray}
    h_i(\boldsymbol \rho) = \alpha \rho_i - \beta \rho_i^2 + \ln (1- \rho_i) - \ln \rho_i
    \label{eq:schelling-utility}
\end{eqnarray}

For simplicity, we define $u(\rho_i) = \alpha \rho_i - \beta \rho_i^2$. Therefore, the potential $\phi(\boldsymbol{\rho})$ in Eq.~\ref{eq:potential} becomes
\begin{eqnarray}
    \phi(\boldsymbol{\rho}) &=& \sum_{i=1}^n \int h_i(\boldsymbol{x}) d x_i = \frac{M}{N}\sum_{i=1}^n \int h_i(\boldsymbol{\rho}) d \rho_i \nonumber\\
    &=&\frac{M}{N}\sum_{i=1}^n \int_0^{\rho_i}\left[ u(\rho) + \ln (1-\rho) - \ln \rho \right] d \rho \nonumber\\
    \label{eq:schelling-potential}
\end{eqnarray}
One can generalize the discrete formulation to a continuous one by replacing $\rho_i$ by $\rho(r)$, where the density is a continuous function of radius $r$ of the neighborhood as demonstrated by Sivaram and Venkatasubramanian~\cite{venkat2022garuds} in the self-organized flocking behavior of birds. \\

Now, according to the theory of potential games~\cite{sandholm2010population}, an arbitrage equilibrium is reached when the potential is maximized. We can determine the equilibrium utility, $h^*$, by the Lagrangian multiplier approach mentioned above (Eq.~\ref{eq:lagrangian}), but there exists a simpler alternative that is more convenient for our purposes here. To analyze the equilibrium behavior, we can take the simpler agent-based perspective and exploit the fact that at equilibrium all agents have the same effective utility, i.e., $h_i = h^*$, for all $i$. In other words, 
\begin{eqnarray}
        \alpha \rho^* - \beta \rho^{*2} + \ln (1-\rho^*) - \ln\rho^* = h^* 
        \label{eq:schelling-utility_equil}
\end{eqnarray}

\begin{figure}[!ht]
\includegraphics[width=0.5\textwidth]{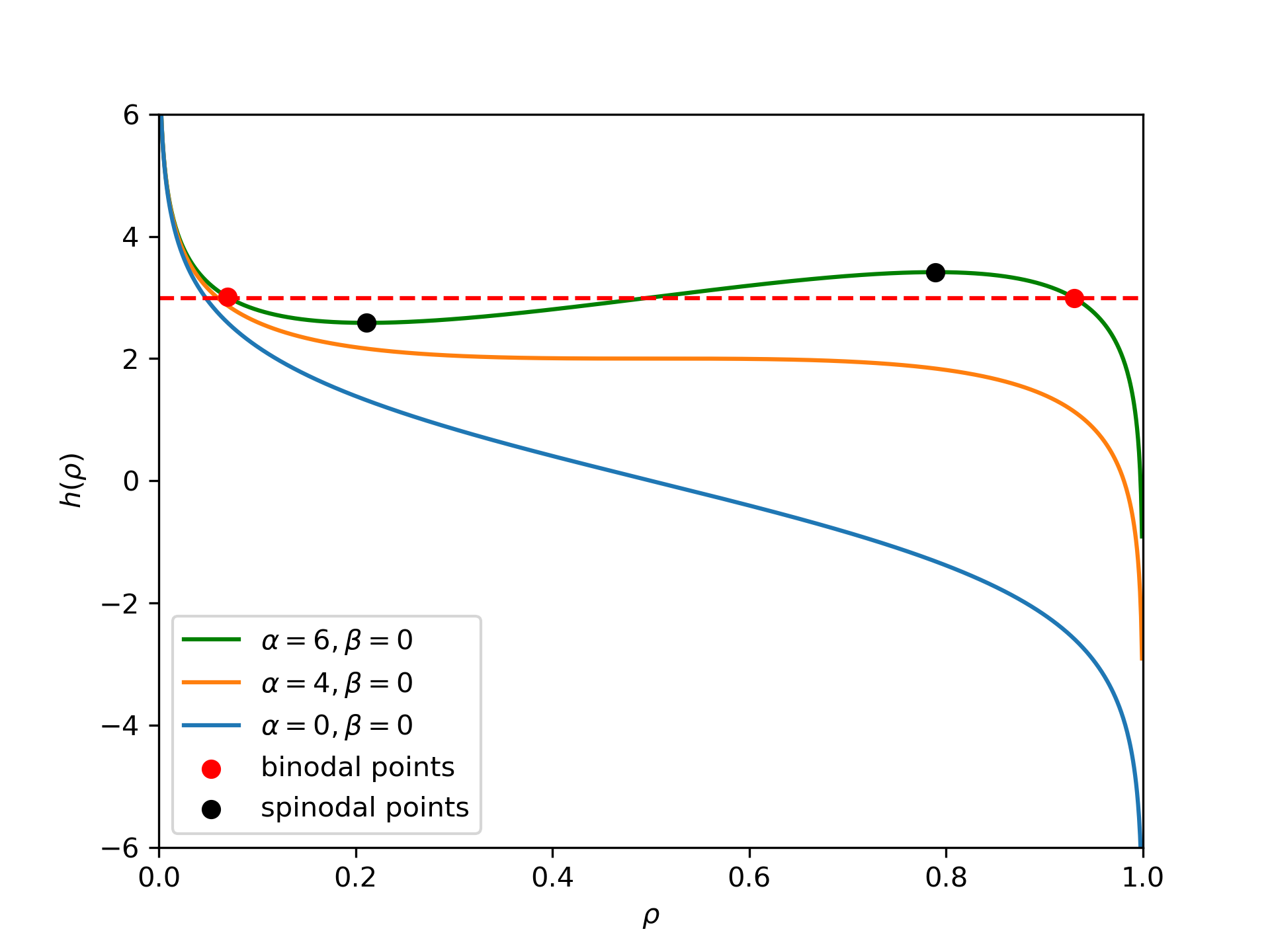}
\caption{Effective Utility vs Density: $h$ vs $\rho$ for different $\alpha$. The black points are the spinodal points ($\rho_{s1} = 0.211, h_{s1} = 2.585; \rho_{s2} = 0.789, h_{s2} = 3.415$). The red points are the binodal points ($\rho_{b1} = 0.071, h_{b1} = 3.00; \rho_{b2} = 0.929, h_{b2} = 3.00$).}
\label{fig:Utility_density_zero_beta}
\end{figure}

\begin{figure}[!ht]
\includegraphics[width=0.5\textwidth]{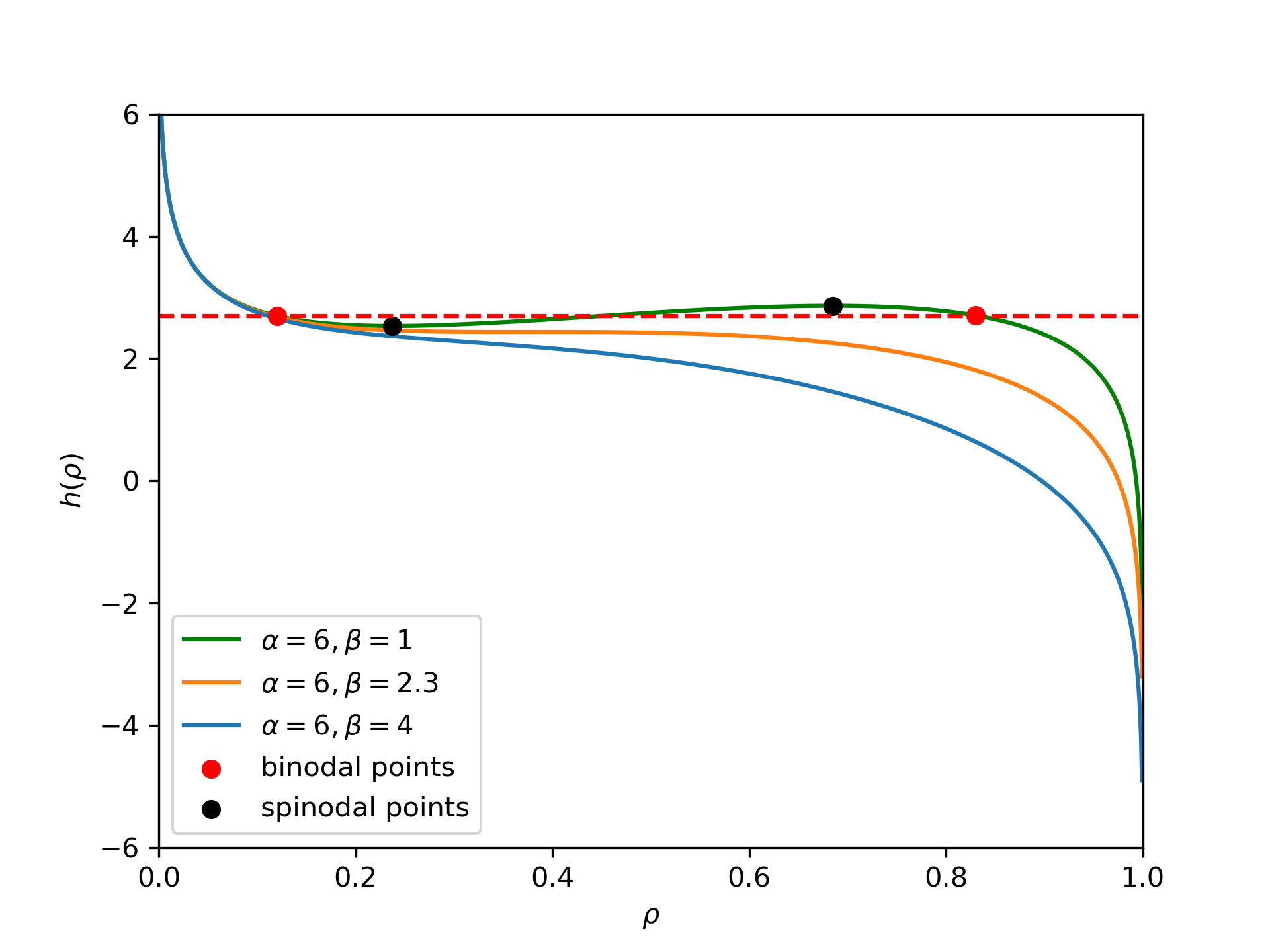}
\caption{Effective Utility vs Density: $h$ vs $\rho$ for different $\beta$
The red points are binodal points. The black points are the spinodal points ($\rho_{s1}$ = 0.237, $h_{s1}$ = 2.535; $\rho_{s2}$ = 0.685, $h_{s2}$ = 2.864). The red points are the binodal points ($\rho_{b1}$ = 0.117, $h_{b1}$ = 2.708; $\rho_{b2}$ = 0.829, $h_{b2}$ = 2.708).}
\label{fig:Utility_density_nonzero_beta}
\end{figure}

We explore numerically the behavior of $h$ as a function of $\rho$ (Eq.~\ref{eq:schelling-utility}), as shown in Fig.~\ref{fig:Utility_density_zero_beta} ($\beta = 0$, different $\alpha$) and Fig.~\ref{fig:Utility_density_nonzero_beta} ($\alpha = 6$, different $\beta$). As we can see, these two plots are qualitatively similar. Below a threshold value of $\alpha$ and $\beta$, the utility function is monotonic and has a unique density (blue curve) for a given utility value. Above the threshold, the utility is non-monotonic (green curve) and can have multiple density values for the same utility. This behavior is known as the \textit{van der Waals loop} in thermodynamics. In mathematical terms, this cubic-like function has three real and positive zeros. The red dotted line shows this. The orange curve represents the threshold behavior. \\

Note that whether all agents remain in a single phase of uniform density dispersed throughout the region or separate into various groups is determined by the slope $\partial h/\partial \rho\big\vert_{\rho^*}$, which is the second derivative of $\phi$, $\partial^2 \phi /\partial^2 \rho\big\vert_{\rho^*}$, and the zeros of $h(\rho)$. This behavior is mathematically equivalent to \textit{spinodal decomposition} in thermodynamics, widely studied, for example, in the phase separation of alloys and polymer blends~\cite{cahn1961spinodal, favvas2008spinodal}. \\

In thermodynamics, the phase between the spinodal points (discussed below in more detail) is unstable because it corresponds to increasing the free energy of the system, and hence the single phase splits into two phases of different densities to lower the free energy.  For the same reason, the phases between the spinodal and binodal points are metastable, and the phases at the binodal points are stable. \\

A similar behavior happens here in statistical teleodynamics as well. Here, the goal is to maximize the potential $\phi$ in \eqref{eq:schelling-potential}. Thus, in Fig~\ref{fig:Utility_density_zero_beta}  and Fig.~\ref{fig:Utility_density_nonzero_beta}, we observe that for $\alpha = 0$ (blue curve) and $\alpha = 4$ (orange curve), $\partial h/\partial \rho \le 0$ (i.e. negative slope; recall that $\partial h/\partial \rho $ = $\partial^2 \phi /\partial^2 \rho $). In such a parameter regime, phase separation does not occur. However, for higher values of $\alpha$, regions with $\partial h/\partial \rho>0$ (i.e., positive slope), phase separation develops. 
\\

\begin{figure}[]
\includegraphics[width=0.5\textwidth]{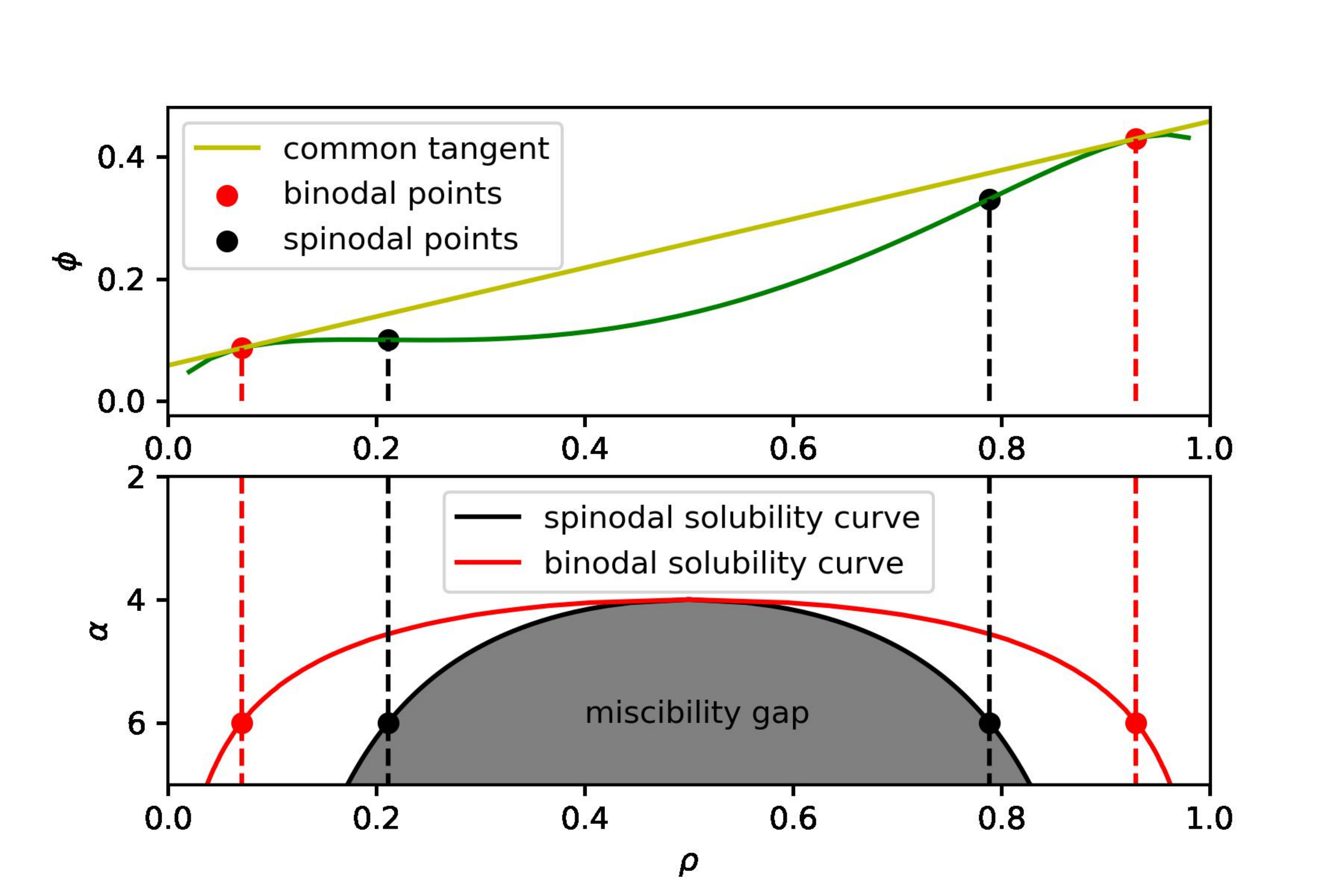}
\caption{Game potential ($\phi$) curve and the spinodal and binodal points. For $\alpha =6, \beta = 0$, the spinodal densities are 0.211 and 0.789; the binodal densities are 0.071 and 0.929.}
\label{fig:SpinodalBinodal}
\end{figure}

We better understand this from Fig.~\ref{fig:SpinodalBinodal}. The upper part of this figure shows the potential ($\phi$) vs the density ($\rho$) curve (in green) for $\alpha =6$, $\beta =0$. 
The plotted equation is
\begin{equation}
    \phi = \alpha \frac{\rho^2}{2} - \beta \frac{\rho^3}{3}- \rho \ln{\rho}- (1-\rho) \ln{(1-\rho)}-2.6 \rho
    \label{eq:schelling-potential-equil}
\end{equation}

The linear term $2.6 \rho$ is subtracted from the actual potential function as a way of rescaling to highlight the \textit{ double hump} nature of the $\phi - \rho$ curve. This subtraction is done purely for illustrative purposes only, as this double hump otherwise is not so visible in the scale of the figure. In all our calculations and simulations, this subtraction is not needed and hence is not done. \\

The spinodal points are shown as black dots, where $\partial h/\partial \rho\big\vert_{\rho^*}$ = $\partial^2 \phi /\partial^2 \rho\big\vert_{\rho^*} = 0$. The corresponding spinodal points are also shown in Fig.~\ref{fig:Utility_density_zero_beta} as black dots on the green curve ($\alpha = 6$, $\beta = 0$). Fig.~\ref{fig:SpinodalBinodal} also shows the binodal points (in red, connected by the common tangent line), where $\partial h/\partial \rho\big\vert_{\rho^*}$ = $\partial^2 \phi /\partial^2 \rho\big\vert_{\rho^*} < 0$. The corresponding binodal points are seen in Fig.~\ref{fig:Utility_density_zero_beta} as red points connected by the red dotted line. As we see, the two binodal points enjoy the same effective utility (3.00), which is the arbitrage equilibrium.\\

The bottom part of Fig.~\ref{fig:SpinodalBinodal} shows the loci of binodal points (red curve) and of spinodal points (black curve) for different values of $\alpha$ ($\beta = 0$). As $\alpha$ changes, the binodal and spinodal points change, and for $\alpha > 4$ ($\beta = 0$) they disappear. Within the spinodal region, shown in dark gray, the miscibility gap, a single phase of uniform density is unstable and would split into two phases of different densities. The reason is that the potential $\phi$ of a large agent group here is less than the sum of the two potentials of the low-density group and the high-density group at the binodal points. We see this geometrically from the common tangent line connecting the binodal points to be above the single-phase green curve between the spinodal points. Agents in such regions will be self-driven towards the high-density binodal point to increase their utility. So $\phi$ increases, and the system splits into two groups of different densities.\\

Thus, for the green curve in Fig.~\ref{fig:Utility_density_zero_beta}, a self-organized, utility-driven, stable phase separation occurs spontaneously at the binodal points (red dotted line) at the arbitrage equilibrium. While the miscibility gap is unstable, the region immediately outside of it, between the black and red curves, is metastable. Beyond the red curve, one has a stable single phase of uniform density - no phase separation here.  \\

In summary, for high values of $\alpha$ (e.g., green curve in Fig.~\ref{fig:Utility_density_zero_beta}), combined with average densities in the miscibility gap, we observe the spontaneous emergence of two phases, high and low density groups of agents, at arbitrage equilibrium, driven by the self-actuated pursuit of maximum utility by the agents. \\

\begin{figure}[!ht]
    \centering
    \includegraphics[width=\linewidth]{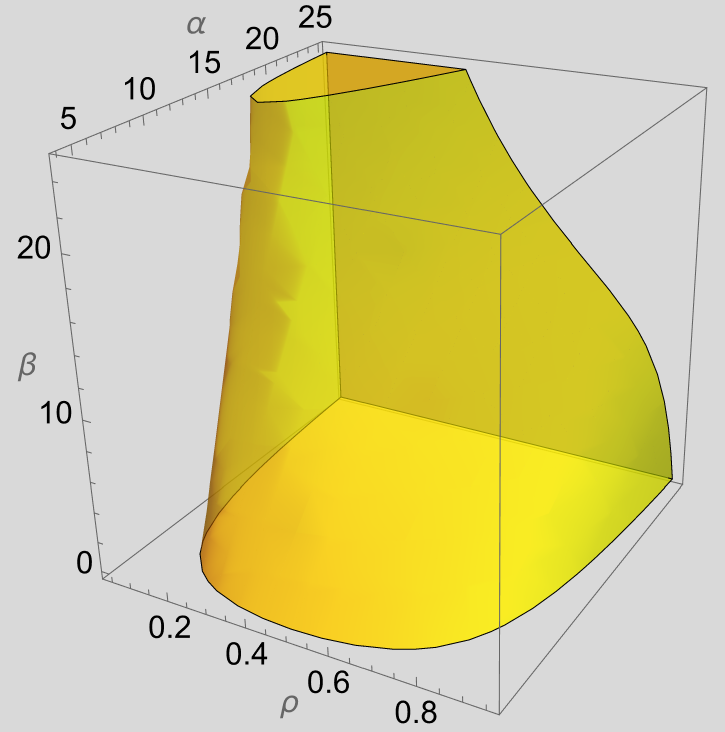}
    \caption{{\bf  Phase separation region at the arbitrage equilibrium}}
    \label{fig:phase_separation}
\end{figure}

\begin{figure}[!ht]
    \centering
    \includegraphics[width=0.5\textwidth]{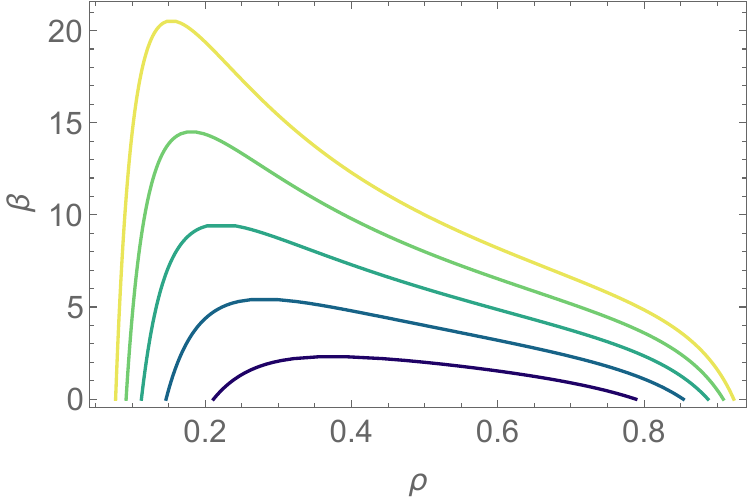}
    \caption{{\bf  Constant $\alpha$ (vertical) 2-D slices of the yellow region}}
    \label{fig:phase_separation_alpha_slice}
\end{figure}

\begin{figure}[!ht]
    \centering
    \includegraphics[width=\linewidth]{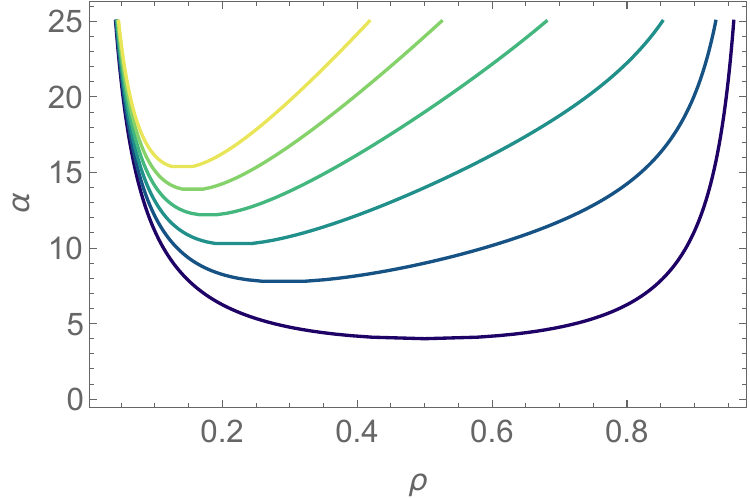}
    \caption{{\bf  Constant $\beta$ (horizontal) 2-D slices of the yellow region}}
    \label{fig:phase_separation_beta_slice}
\end{figure}

In Fig.~\ref{fig:phase_separation}, we show the region (shaded in yellow) within which the phase separation occurs at the arbitrage equilibrium for the values of the average density $\rho_0$, $\alpha$, and $\beta$ within the region. In Fig.~\ref{fig:phase_separation_alpha_slice} and Fig.~\ref{fig:phase_separation_beta_slice}, we show the 2-D slices of the yellow region of spontaneous phase separation. For a given value of $\alpha$, $\beta$, and $\rho_0$, they show the loci of the two densities (i.e., low- and high-density groups) of the corresponding equilibrium states of the agents.\\

Intuitively, in the high-density phase, agents derive so much more benefit from the affinity term (due to the high $\alpha$) that it more than compensates for the disutilities due to congestion and competition, thus yielding a high effective utility. Similarly, in the low-density phase, the benefits of reduced congestion and lower competition combined with increased option benefit more than compensate for the loss of utility from the affinity term. Thus, every agent enjoys the same effective utility $h^*$ in one phase or the other at the arbitrage equilibrium. This causes equilibrium because, as noted, there is no more arbitrage incentive left for agents to switch neighborhoods.  \\

As noted, this analysis is mathematically equivalent to spinodal decomposition in statistical thermodynamics, with an important difference. In statistical thermodynamics, agents try to \textit{minimize} their chemical potentials and the free energy of the system. Here, in statistical teleodynamics, agents try to \textit{maximize} their utilities ($h_i$) and the game-theoretic potential ($\phi$). In thermodynamics, chemical potentials are equal in phase equilibrium. In teleodynamics, the effective utilities are equal in arbitrage equilibrium.\\

The parallel is striking, but it is not surprising because, as Venkatasubramanian has shown~\cite{venkat2017book, venkat2022unified}, statistical teleodynamics is a generalization of statistical thermodynamics for goal-driven agents. Therefore, given this mathematical equivalence, one should expect to observe ``macroscopic" phenomena generally associated with thermodynamics (such as phase separation and equilibrium) in entirely different contexts (such as MIPS or social segregation in socioeconomic systems). The ``microscopic" mechanisms of the self-organizing dynamics might differ in different contexts. Thus, the driving force for the movements of nonliving agents could be temperature, pressure, or chemical potential gradients, whereas the driving force for living agents is effective utility. As noted, since our theory is ``mesoscopic" in character, it is agnostic to the ``microscopic" details. 

\subsection{Agent-based simulation results}

We tested our model using an agent-based simulation on a $300\times 300$ lattice (90,000 cells total), for $\alpha = 6$, $\beta = 0$, and for different $N$ ($N$ = 22,500, 45,000, 55,000). For details of the simulations, the reader is referred to the Methods section. \\
\begin{figure*}[!ht]
\centering
\includegraphics[scale=0.4]{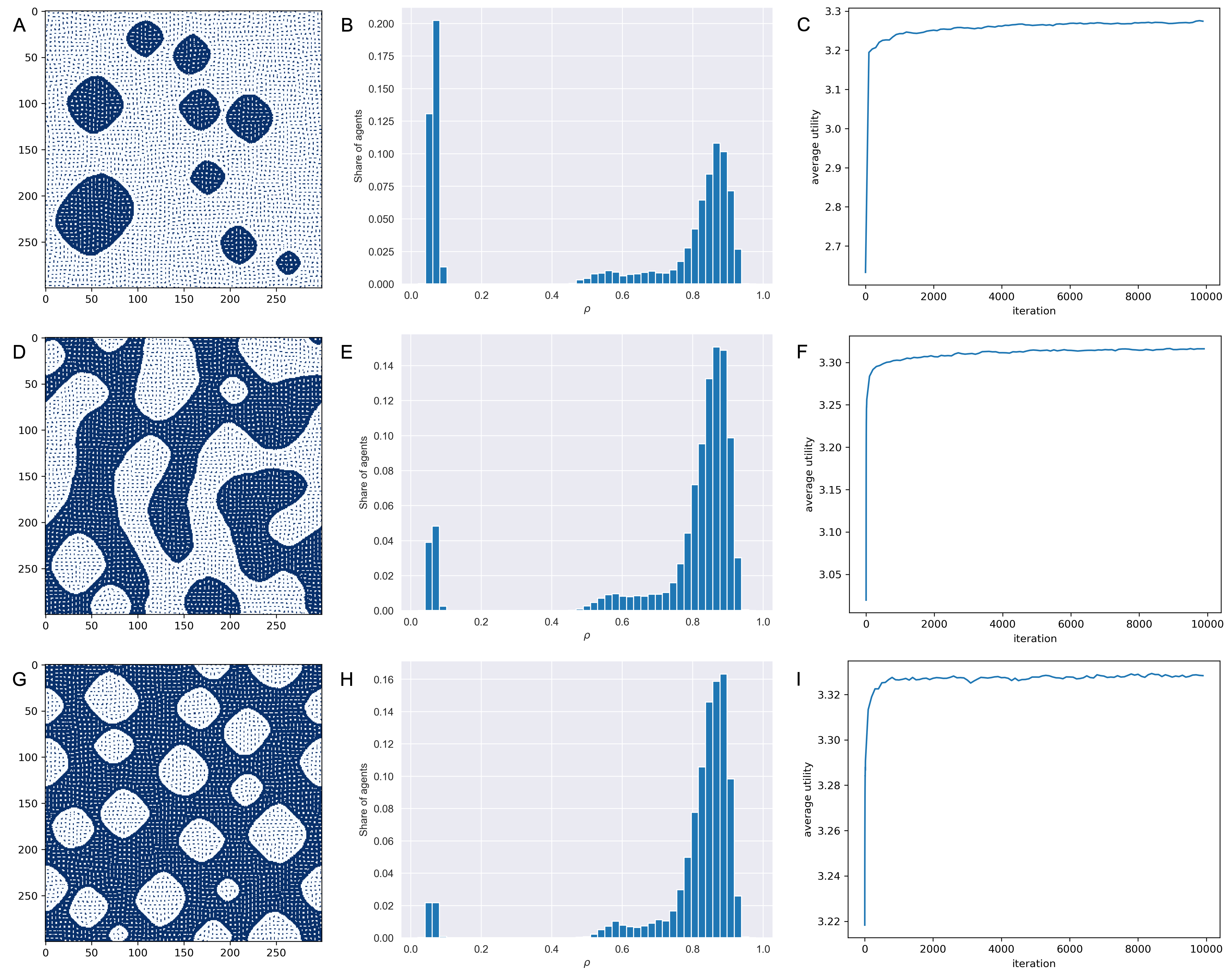}
\caption{Equilibrium patterns for different average densities, $\rho_0$, at the end of 10,000 iterations, for $\alpha = 6,$ $\beta = 0$.
(A), (B), and (C) represent, respectively, the sparsely distributed dots pattern (I) obtained for 22,500 agents ($\rho_0 = 0.25$), the corresponding density distribution at the end of 10,000 iterations, and the average utility evolution over the iterations. Similar results are shown for the labyrinthine pattern (II) in (D), (E), and (F) for 45,000 agents ($\rho_0 = 0.5 $), and for the ``gapped" pattern (III) in (G), (H), and (I) for 55,000 agents ($\rho_0 =0.61 $).} 
\label{fig:agentPatterns_3X3}
\end{figure*}

In our simulations, we observe (Fig.~\ref{fig:agentPatterns_3X3}) the three basic types of patterns, or ``macroscopic" states, namely, (I) sparsely distributed dots, (II) labyrinthine or worm-like structures, and (III) ``gapped" patterns  that are seen empirically, for example, in mussel beds~\cite{liu2013CahnHilliard} for different mussel densities. As one might expect, the size of the interaction neighborhood (see the Methods section) plays a role in determining the specific details, i.e., the ``microscopic" features, of these ``macroscopic" states. That is, for example, the detailed ``microscopic" features of the labyrinthine or worm-like structures might look different for different neighborhood sizes, but their ``macroscopic" state would remain labyrinthine. Thus, the basic ``macroscopic" states are found to be robust. The ``macroscopic" state transitions appear like phase transitions, moving from category I to category II to III as the density increases. \\

We believe that these ``macroscopic" and ``microscopic" characteristics reflect the structure of the phase-space landscape of $\phi$. Thus, as the agents move around in the physical space, the system wanders around in the phase-space landscape, settling into one state or the other. Since a ``macroscopic" state could be achieved via many different ``microscopic" states, i.e., \emph{multiplicity}, one gets different ``microscopic" outcomes in different simulation runs, while the ``macroscopic" outcome remains the same and robust. The ``macroscopic" states are the attractors seen in many nonlinear dynamical systems. \\

The corresponding density histograms and the evolution of average utility as a function of iteration are also shown in Fig.~\ref{fig:agentPatterns_3X3}. For this configuration (i.e., $\alpha = 6$, $\beta = 0$), the spinodal densities (the black points in Fig.~\ref{fig:Utility_density_zero_beta} and Fig.~\ref{fig:SpinodalBinodal}) are 0.211 and 0.789, and the binodal densities (the red points in Fig.~\ref{fig:Utility_density_zero_beta} and Fig.~\ref{fig:SpinodalBinodal}) are 0.071 and 0.929. \\

A word of caution as we discuss the results. As noted, our simple coarse-grained model is equivalent in spirit to the van der Waals or the Ising model in statistical mechanics. Therefore, we do not expect our analytical model and associated simulations to capture all the nuances and complexities of the real-world patterns in active matter systems. \\

As predicted by our theory, the density histograms show (Fig.~\ref{fig:agentPatterns_3X3}, B, E, H) spinodal decomposition for the three agent populations ($N$ = 22,500, 45,000, 55,000). That is, there are two phases, one with low-density groups and the other with high-density groups. These two densities are around the binodal densities predicted by the theory. Although the theory predicts two sharp binodal density values ($\rho_1$ = 0.071 and $\rho_2$ = 0.929), it would be hard to see such precise results in the simulation for one main reason. Theoretical predictions are based on concepts from statistical mechanics, which only work well for an extremely large number of agents (such as the Avogadro number of molecules, $\sim10^{23}$). This is when the statistical estimates and outcomes are extremely precise, as in the case of, for example, alloys in materials science. In our simulation, we have only 22,500 - 55,500 agents. So, the statistics is not that precise. Therefore, one should expect to see a distribution of values instead of singular peaks. That is what we observe in our simulations. \\

We also observe that the distributions around the low binodal density are narrower, whereas they are broader at the high binodal density. The reason is the following. As we see in Fig.~\ref{fig:Utility_density_zero_beta}, an individual agent reaches its maximum utility at the upper spinodal point at the spinodal density of 0.789, while the entire agent bed reaches its maximum potential $\phi$ (and hence the arbitrage equilibrium) at the binodal density of 0.929 as seen from Fig.~\ref{fig:SpinodalBinodal}. Thus, in high-density clusters, individual agents constantly compete to reach the upper spinodal point (density = 0.789) of higher individual utility, while the competition of the other agents to reach the same state drives the agent bed away from the spinodal point to the binodal point (density = 0.929). Therefore, agents mainly bounce around between these two points, the spinodal density of 0.789 and the binodal density of 0.929, with a weighted average density of about 0.85 (see Table~\ref{tab:summary}) right in the middle. \\

We also observe in Fig.~\ref{fig:agentPatterns_3X3} (C, F, I) that the average utility improves as the simulation proceeds, as the agents maneuver around to increase their effective utilities, and then finally settles and fluctuates around the arbitrage equilibrium value.  \\

The key statistics are summarized in Table~\ref{tab:summary}. The spinodal and binodal densities are the same for the three different cases of $N$, because $\alpha = 6, \beta =0$ for all cases (see Fig.~\ref{fig:Utility_density_zero_beta}, green curve). We also find that the average utility of Phase-1 is almost the same as that of the corresponding Phase-2, as predicted by the theory. Thus, we see that a vast majority (86-90\%) of the agents are in their arbitrage equilibrium states, either in Phase-1 or Phase-2.

\begin{table*}[]
  \centering
  \caption{Summary of key metrics in phase separation}
    \label{tab:summary}
    \begin{adjustbox}{width=\linewidth,center}
  \begin{tabular}{c c c c c c c c c c c c}
    \toprule
    \multirow{2}{1cm}{Category} & \multirow{2}{1cm}{N} & \multicolumn{2}{c}{Spinodal densities}  & \multicolumn{2}{c}{Binodal densities} & \multicolumn{3}{c}{Phase-1} & \multicolumn{3}{c}{Phase-2}  \\ 
    
        {}                     & {}                      &   {$\rho_{s1}$} & $\rho_{s2}$      & $\rho_{b1}$ & $\rho_{b2}$       & \multirow{2}{2cm}{Share of agents($\%$)} & Avg. density & Avg. utility   & \multirow{2}{2cm}{Share of agents($\%$)} & Avg. density & Avg. utility   \\

        {}   & {} & {}   & {} &{}   & {} &{}   & {} &{}   & {} &{}   & {} \\
        \hline
        {Sparsely distributed dots}   & {22,500} & {0.211}   & {0.789} &{0.071}   & {0.929} &{33.28}   & {0.052} &{3.217}   & {52.63} &{0.855}   & {3.330} \\
        {Labyrinthine}   & {45,000} & {0.211}   & {0.789} &{0.071}   & {0.929} &{8.71}   & {0.051} &{3.234}   & {79.85} &{0.850}   & {3.338} \\
        {Gapped}   & {55,000} & {0.211}   & {0.789} &{0.071}   & {0.929} &{4.32}   & {0.050} &{3.250}   & {85.42} &{0.848}   & {3.341} \\

    \hline
  \end{tabular}
\end{adjustbox}
\end{table*}

\subsection{Stability of the Arbitrage Equilibrium}

We can determine the stability of this equilibrium by performing a Lyapunov stability analysis \cite{venkat2015howmuch, venkat2017book}. A Lyapunov function $V$ is a continuously differentiable function that
takes positive values everywhere except at the equilibrium point (i.e., $V$ is positive definite), and decreases (or is not increasing) along every
trajectory traversed by the dynamical system ($\dot{V}$ is negative definite or negative semi-definite). A dynamical system is locally stable at equilibrium if $\dot{V}$ is negative semi-definite and is asymptotically stable if $\dot{V}$ is negative definite.\\

Following Venkatasubramanian~\cite{venkat2017book}, we identify our Lyapunov function $V(\boldsymbol{\rho})$

\begin{eqnarray}
V(\boldsymbol{\rho}) = \phi^*(\boldsymbol{\rho}) - \phi(\boldsymbol{\rho}) \label{eq:lyapunov}
\end{eqnarray}

where $\phi^*$ is the potential at the arbitrage equilibrium (AE) (recall that $\phi^*$ is at its maximum at AE) and $\phi(\boldsymbol{\rho})$ is the potential at any other state. Note that $V(\boldsymbol{\rho})$ has the desirable properties we seek: (i) $V(\boldsymbol{\rho^*})$ = 0 at AE and $V(\boldsymbol{\rho})$ $>$ 0 elsewhere, i.e., $V(\boldsymbol{\rho})$ is positive definite; (ii) Since $\phi(\boldsymbol{\rho})$ increases as it approaches the maximum, $V(\boldsymbol{\rho})$ decreases with time, so it is easy to see that $\dot{V}$ is negative definite. Therefore, the arbitrage equilibrium is not only stable but also \textit{asymptotically stable}. \\

\begin{figure*}[!ht]
    \centering
    \includegraphics[width=\linewidth]{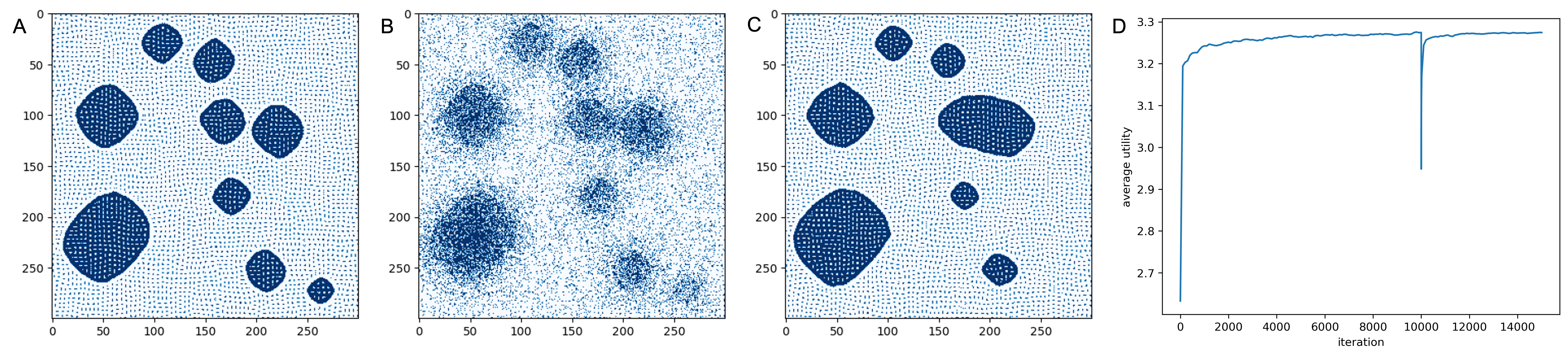}
    \caption{{\bf  Stability analysis. (A) Equilibrium configuration of agents at the end of 10,000 iterations (B) disturbed configuration at 10,001$^\text{th}$ iteration (C) equilibrium configuration at the end of 15,000 iteration  (D) evolution of average utility over iterations. The sharp decrease in the average utility at 10,001$^\text{th}$ is due to the disturbance introduced at 10,001$^\text{th}$ iteration.}}
    \label{fig:Stability_analysis_agent_locations}
\end{figure*}

Our simulation results confirm this theoretical prediction (see Fig.~\ref{fig:Stability_analysis_agent_locations}). We show the stability results for the configuration of Fig.~\ref{fig:agentPatterns_3X3}-A, as an example. After the agents population reached equilibrium (10,000 iterations), we disturbed the equilibrium state by randomly changing the positions of the agents. As a result, the average utility of the population goes down as seen from the sharp drop at the 10,001th iteration in Fig.~\ref{fig:Stability_analysis_agent_locations}-D. Fig.~\ref{fig:Stability_analysis_agent_locations}-B shows the disturbed state of the agent groups. The simulation is then continued from this new disturbed far-from-equilibrium state. As we see from Fig.~\ref{fig:Stability_analysis_agent_locations}-C, the agent population recovers quickly to reach the original category-I ``macroscopic`` state even though some of the ``microscopic" features are different this time. The reader might have noticed that two small groups in Fig.~\ref{fig:Stability_analysis_agent_locations}-A  have merged to become one larger group in two different locations in Fig.~\ref{fig:Stability_analysis_agent_locations}-C. This phenomenon is called Ostwald Ripening in materials science. We also notice that the average utility is back to its old level. \\

This analysis shows that the arbitrage equilibrium region is not only stable, but asymptotically stable. That is, the ``macroscopic" structures are resilient and self-healing. Given the speed of recovery, it could possibly be exponentially stable, but we have not proved this analytically here. It is interesting to observe that this result is similar to that of the dynamics of the income game~\cite{venkat2015howmuch, venkat2017book} and the dynamics of flocking birds~\cite{venkat2022garuds}. 

\section{Connection with Statistical Thermodynamics}

To appreciate how statistical teleodynamics is similar to and different from statistical thermodynamics, let us consider a familiar example, the thermodynamic system of gas molecules in a container, from the perspective of statistical teleodynamics. We call this the \textit{Thermodynamic Game}~\cite{venkat2015howmuch, venkat2017book}. 
Now, real gas molecules are, of course, purpose-free, and hence do not pursue maximum utility. However, in this game-theoretic formulation, we show that our imaginary molecules-like agents, when they pursue a particular form of ``utility" as given in Eq.~\ref{eq:thermo_utility}, behave like gas molecules. So, approaching this from the perspective of a potential game, we introduce the following ``utility," $h_i$, for our molecular agents in state $i$:
\begin{equation}
{h}_i(E_i,N_i)=-\beta E_i - \ln N_i
\label{eq:thermo_utility}
\end{equation}

where $E_i$ is the energy of an agent in state $i$, $N_i$ is the number of agents in state $i$, $\beta = 1/k_BT$, $k_B$ is the Boltzmann constant and $T$ is temperature. The first term models the tendency of molecules to prefer low-energy states (since our ``molecular" agent tries to maximize its utility, the negative sign leads to smaller values of $E_i$). The $-\ln N_i$ term models the \textit{disutility of competition} we have used in the sections above. This term models the ``restless" nature of molecules and their propensity to spread out. This is so because the $-\ln N_i$ term incentivizes the agent to leave its current location of higher $N_i$ to a location of lower $N_i$ all the time. \\ 

By integrating this effective utility, we obtain the potential $\phi(\mathbf{x})$ as
\begin{equation}
\phi(\mathbf{x})=-\frac{\beta}{N} E +\frac{1}{N} \ln\frac{N!}{\prod_{i=1}^n(Nx_i)!}\label{eq:thermo_potential}
\end{equation}

where $E=N\sum_{i=1}^nx_iE_i$ is the total energy that is conserved, $N$ is the total number of molecules, $n$ is the total number of states, and $x_i = N_i/N$.\\

This game reaches a unique Nash equilibrium when $\phi(\mathbf{x})$ is maximized~\cite{sandholm2010population}. To determine the equilibrium distribution, we maximize the Lagrangian given by Eq.~\ref{eq:lagrangian}
\begin{equation*}
L=\phi+\lambda(1-\sum_{i=1}^nx_i)
\end{equation*}

and obtain the well-known Boltzmann exponential distribution of energy at equilibrium 
\begin{equation}
x_i^*=\frac{\exp(-\beta E_i)}{\sum_{j=1}^n\exp(-\beta E_j)}\label{eq:boltzmann}
\end{equation}

Thus, the arbitrage equilibrium of this game is the same as the statistical thermodynamic equilibrium, as expected. The critical insight here is, as Venkatasubramanian et al. first showed~\cite{venkat2015howmuch}, that the second term in Eq.~\ref{eq:thermo_potential} is the same as entropy (except for the Boltzmann constant). Thus, maximizing potential in population game theory is equivalent to maximizing entropy in statistical mechanics subject to the constraints given by the first term in Eq.~\ref{eq:thermo_potential}, i.e., the constraint on total energy $E$. \\

This is a deep and beautiful connection between statistical mechanics and potential game theory. This fundamental connection allows us to generalize statistical thermodynamics to statistical teleodynamics and lays the foundation towards a universal theory of emergent equilibrium behavior of both purpose-free and purpose-driven agents.\\ 

Pursuing this line of enquiry some more, we recognize another important connection. From Eq.~\ref{eq:thermo_potential}, we have
\begin{equation}
\phi=-\frac{1}{Nk_BT} (E-TS)=-\frac{\beta}{N} A
\label{eq:thermo-helmholtz}
\end{equation}

where $A = E - TS$ is the Helmholtz free energy. Indeed, in statistical thermodynamics, $A$ is called a {\em thermodynamic potential}. Again, we see the correspondence between the game-theoretic potential and the thermodynamic potential.  Furthermore, we see the correspondence between utility and the {\em chemical potential}, with an important difference. Active agents try to increase their utilities, whereas thermodynamic agents try to decrease their chemical potential.  In the \textit{Thermodynamic Game}, an arbitrage equilibrium is reached when all agents have the same effective utility, which is equivalent to the chemical potential being equal in thermodynamics. In fact, our theory reveals the critical insight that both living and non-living agents are driven by arbitrage opportunities towards equilibrium, except that their \textit{arbitrage currencies} are different. For thermodynamic agents, the currency is the chemical potential, whereas for living agents, the utility. Thus, we see that statistical teleodynamics is a natural generalization of statistical thermodynamics for goal-driven agents.\\

The perspective of statistical teleodynamics reveals an underappreciated insight in statistical mechanics, which is the recognition that when $N_i$ (or $x_i$) follows the exponential distribution in Eq.~\ref{eq:boltzmann}, $h_i = h^*$ for all molecules. In other words, it is an \textit{invariant}. As noted, this is the defining criterion for arbitrage equilibrium. Thus, the exponential distribution is a curve of \textit{constant effective utility} (or equivalently constant chemical potential), for all values of $E_i$. That is, it is an \textit{ \textit{isoutility}} curve for $h_i$ defined by Eq.~\ref{eq:thermo_utility}. This turns out to be a particularly valuable insight in the context of the emergence of a fair income distribution in the dynamics of the free market~\cite{venkat2017book}. \\

This recognition reveals another valuable insight that is also not readily appreciated in statistical mechanics. We realize that we do not necessarily have to maximize the potential $\phi(x)$ to derive the equilibrium distribution (when the equilibrium is unique, i.e., $\partial^2 \phi /\partial^2 x < 0$). We can adopt the \textit{agent} perspective and recognize that equilibrium is reached when all agents enjoy the same utility $h_i = h^*$. Therefore, we have 

\begin{equation}
{h}_i= {h^*} = -\beta E_i - \ln N_i^*, ~~i \in \{1,\dots,n\}
\label{eq:equil-h}
\end{equation}

where $N_i^*$ is given by the equilibrium distribution. From this, it is easy to derive the Boltzmann energy distribution (Eq.~\ref{eq:boltzmann}) by rearranging and solving for $N_i^*$. In fact, this is the trick we used to derive Eq.~\ref{eq:ant-x_i} from Eq.~\ref{eq:ant-utility-equil}. \\

Thus, we see that without necessarily invoking the system perspective of maximizing the potential (which is equivalent to maximizing entropy with constraints), we can derive the Boltzmann distribution easily. This important property is not seen that clearly in statistical mechanics. In statistical mechanics, we typically maximize entropy or minimize Gibbs free energy to arrive at equilibrium results. That is, the emphasis is on the \textit{system} perspective; the individual agent's view is not given importance. This is one of the important philosophical differences between statistical thermodynamics and statistical teleodynamics. While the former is generally a top-down, system-oriented perspective, the latter is decidedly a \textit{bottom-up, agent-oriented} 
 perspective. \\

Both Eq.~\ref{eq:thermo-helmholtz} and Eq.~\ref{eq:thermo_utility} reveal another interesting feature of the statistical teleodynamics framework with respect to thermodynamic laws~\cite{venkat2004spontaneous, venkat2006entropy, venkat2007darwin}. Since maximizing potential $\phi$ is the same as maximizing entropy $S$ subject to the constraint on $E$ (Eq.~\ref{eq:thermo-helmholtz}), we see the two laws of thermodynamics embedded in this equation. The second term is the root of the second law of maximizing entropy and the first term is the source of the first law of energy conservation. Similarly, in Eq.~\ref{eq:thermo_utility}, the two laws are embedded in the same manner. 
Eq.~\ref{eq:thermo-helmholtz} is the \textit{system-based} view of the two laws, while Eq.~\ref{eq:thermo_utility} is the \textit{agent-based} view. \\

Therefore, we realize that the non-entropic terms in utility and potential serve the role of \textit{constraints} on entropy maximization. This interpretation provides a deeper understanding of the effective utility functions (and their potential versions) of living agents, such as ants. Consider the effective utility of an ant given by Eq.~\ref{eq:ant-utility}. The $-\ln N_i$ term, as before, corresponds to the second law of entropy maximization, and the other two terms play the role of the first law of enforcing constraints. However, unlike the thermodynamics first law of energy conservation, the teleodynamics ``first law" does not enforce conservation, but rather constraints. Thus, we see that the teleodynamics ``first law" is the general case, and the thermodynamic first law is a special case where the constraint on the total energy is also the conservation of it. The ``first law" of teleodynamics allows for more complicated and ``non-thermodynamic" constraints to be imposed on entropy maximization. We also see that the ``second law" of teleodynamics is similar to the second law of thermodynamics, but with the broader concept of arbitrage equilibrium. The ''zeroth law" of teleodynamics is that all agents continually try to increase their effective utilities. This is essentially the Darwinian survival-of-the-fittest principle in biology and the Smithian constant pursuit of self-interest principle in economics. 

\section{Connection with MIPS}

It is instructive to compare the phase separation phenomena described by our analysis of \textit{non-thermodynamic} systems, such as mussel beds~\cite{venkat2023mussels} or social groups~\cite{venkat2024social}, with MIPS. MIPS is generally described as a non-equilibrium or out-of-equilibrium behavior of active matter systems~\cite{cates2013active, cates2015motility, brady2015thermoActivematter, brady2023MechnicalMIPS,ramaswamy2010mechanics, marchetti2013hydrodynamics}. Based on our analysis above, we argue that MIPS is indeed an equilibrium phenomenon, but a different kind of equilibrium, an \textit{arbitrage equilibrium}. As we show above, under certain conditions, this arbitrage equilibrium is equivalent to a statistical or thermodynamic equilibrium. In this section, we explore this perspective at some length. \\

We start with the work reported by Takatori and Brady~\cite{brady2015thermoActivematter}. In particular, four of their equations are relevant to our discussion here. They are for the active pressure ($\mathrm{\Pi^{act}}$), Helmholtz free energy ($\mathrm{F^{act}}$), Gibbs free energy ($\mathrm{G ^{act}}$), and the chemical potential ($\mathrm{\mu^{act}}$) of the self-propelled particles, respectively, as given below (we have changed their volume fraction symbol $\phi$ to $\theta$ so that it is not confused with our $\phi$ which is the game-theoretic potential).
\begin{eqnarray}
  \Pi^{act} &=& n k_s T_s [1 - \theta - \theta ^2 + 3 \theta \mathrm{Pe_R} (1 - \theta/\theta_0)^{-1}]
    \label{eq:pi-act}
\end{eqnarray}
\begin{eqnarray}
    \frac{F^{act}}{N k_s T_s}  = \ln \theta - \frac{\theta (\theta + 2)}{2} - 3 \mathrm{Pe_R} \theta_0 \ln (1 - \theta/\theta_0) \nonumber \\  + F^{0}(k_sT_s, \mathrm{Pe_R})
    \label{eq:F-act}    
\end{eqnarray}
\begin{eqnarray}
     \frac{G^{act}}{N k_s T_s}  &=& \frac{F^{act}}{N k_s T_s} + \frac{\Pi^{act}}{n k_s T_s} 
    \label{eq:G-act}
\end{eqnarray}
\begin{eqnarray}
    \mu^{act}(k_s T_s, \theta, \mathrm{Pe_R})  = \mu^{0}(k_s T_s, \mathrm{Pe_R}) + k_s T_s \ln \theta 
    \nonumber \\ + k_s T_s \ln \Gamma (\theta, \mathrm{Pe_R})
    \label{eq:mu-act}    
\end{eqnarray}

where $\theta$ is the volume fraction, $\theta_0$ is the volume fraction at close packing, $N$ is the number of active swimmers, $k_sT_s = \zeta U_0^2 \tau_R/6$, $U_0$ is the intrinsic swim speed, $\tau_R$ is the reorientation time, $\zeta$ is the hydrodynamic drag factor, $\mathrm{Pe_R}$ is the Peclet number, $\mathrm{\Pi^{act}}$ is the active pressure, $\mathrm{F^{act}}$ is the non-equilibrium Helmholtz free energy, $\mathrm{G^{act}}$ is the nonequilibrium Gibbs free energy, $\mathrm{F^0}$ is the reference state Helmholtz free energy, $\mu^{act}$ is the nonequilibrium chemical potential, $\mu^{0}$ is the reference state of the chemical potential, $\Gamma$ is a nonlinear expression (for more details on these quantities,  see~\cite{brady2015thermoActivematter}).\\

In Eq.~\ref{eq:mu-act}, the second term on the right-hand side represents the
entropic, ideal-gas contribution to the chemical potential. The
third term is the nonideal term that is the analog of enthalpic attraction between active swimmers, and is represented by $\Gamma (\theta, Pe_R)$ that resembles the fugacity coefficient in classical thermodynamics. \\

We observe that our Eq.~\ref{eq:schelling-potential-equil} is equivalent to Eq.~\ref{eq:G-act}. The game-theoretic potential is the negative of the Gibbs free energy, since our active agents maximize utility rather than minimizing the chemical potential. Our Fig.~\ref{fig:SpinodalBinodal} showing the spinodal and binodal regions is equivalent to Figure 3 in their paper (except for sign reversal due to game potential $\phi = - G^{act} $). Note that we are not saying that they are equal; they are equivalent in the sense that both produce a function ($\phi$ or $G^{act}$) with two maxima (for $\phi$) or equivalently two minima (for $G^{act}$) so that the common tangent line can be drawn to determine the binodal points.  Thus, they are two equivalent models of phase separation in active matter systems. \\

Now, from Eq.~\ref{eq:schelling-utility-non-simplified} ($\delta = 1$), we have
\begin{eqnarray}
    h_i(\boldsymbol \rho) = \alpha \rho_i - \beta \rho_i^2 + \gamma \ln (1- \rho_i/\rho_{\mathrm{max}}) - \ln \rho_i
    \label{eq:schelling-utility-rho-max}
\end{eqnarray}

The nonequilibrium chemical potential $\mu^{act}$ (Eq.~\ref{eq:mu-act}) is equivalent to our effective utility $h_i$ in Eq.~\ref{eq:schelling-utility-rho-max}. We see that our entropic term $-\ln \rho_i$ corresponds to their $k_s T_s \ln \theta$ (the signs are opposite because the utility is negative of the chemical potential), and the rest are reflected in $\ln \Gamma (\theta, \mathrm{Pe_R})$, where $\Gamma (\theta, \mathrm{Pe_R})$ is given by~\cite{brady2015thermoActivematter}
\begin{eqnarray}
\Gamma (\theta, \mathrm{Pe_R}) = (1- \theta/\theta_0)^{-3\theta_0 \mathrm{Pe_R}} \exp[\theta^3 - \theta^2/2 \nonumber \\+ 3\mathrm{Pe_R}\theta_0 (1- \theta_0)/(1-\theta/\theta_0)
\nonumber \\- 3\theta(1 - \theta_0 \mathrm{Pe_R})]
\end{eqnarray}
and therefore
\begin{eqnarray}
\ln \Gamma (\theta, \mathrm{Pe_R}) =-3\mathrm{Pe_R} \theta_0 \ln (1- \theta/\theta_0) 
+(\theta^3 - \theta^2/2)
\nonumber \\+ 3\mathrm{Pe_R}\theta_0 (1- \theta_0)/(1-\theta/\theta_0)
\nonumber \\- 3\theta(1 - \theta_0 \mathrm{Pe_R}) \nonumber \\
\end{eqnarray}

\begin{figure}[!ht]
    \centering
    \includegraphics[width=0.5\textwidth]{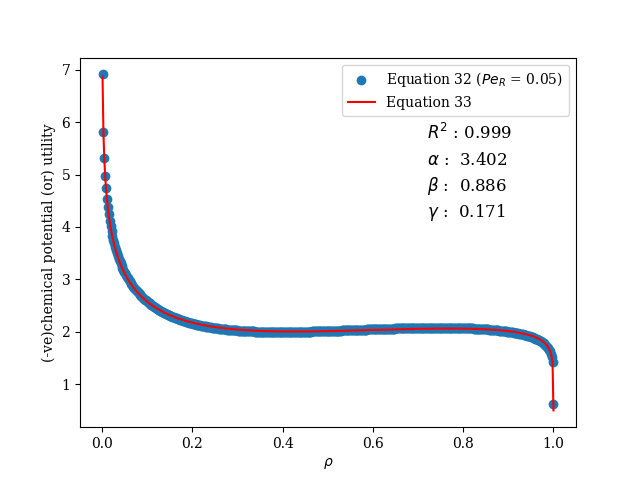}
    \caption{{\bf Effective utility vs Density} with $\theta_0 = 1$ and $\mathrm{Pe_{R}} = 0.05$}
    \label{fig:Brady-chem-poten}
\end{figure}

Their $\ln \Gamma (\theta, \mathrm{Pe_R})$ is much more complicated than our equivalent terms, as it captures all the detailed phenomenology of the active particle dynamics, whereas our model, again, is agnostic of such details. We show in Fig.~\ref{fig:Brady-chem-poten} the fit of Eq.~\ref{eq:schelling-utility-rho-max} to the data from Eq.~\ref{eq:mu-act}, for $\theta_0 = 1$ ($\rho_{\mathrm{max}} = 1$) and $Pe_\mathrm{R} = 0.05$, as an example. We see that despite ignoring the complicated details in Eq.~\ref{eq:mu-act}, the simpler Eq.~\ref{eq:schelling-utility-rho-max} fits the true $\mu^{act}$ quite well. \\

The purpose of this exercise is not to accurately predict Eq.~\ref{eq:mu-act} using Eq.~\ref{eq:schelling-utility-rho-max}. Obviously, Eq.~\ref{eq:mu-act} will always do better than Eq.~\ref{eq:schelling-utility-rho-max} as it incorporates more of the ``microscopic" details of the dynamics that Eq.~\ref{eq:schelling-utility-rho-max} ignores. The objective is to show that the simpler ``mesoscopic" generic model in Eq.~\ref{eq:schelling-utility-rho-max}, which can be used as a template for MIPS in different domains including non-physicochemical phenomena such as social segregation~\cite{venkat2024social}, has sufficient predictive and explanatory power to do just as well as the more detailed customized model in Eq.~\ref{eq:mu-act}. \\

Therefore, we argue that what Takatori and Brady describe as ``nonequilibrium chemical potential" and ``free energy of nonequilibrium active matter" in their paper are actually equilibrium quantities, namely, \textit{arbitrage equilibrium} quantities. As our analysis above (Figures~\ref{fig:SpinodalBinodal}-\ref{fig:phase_separation_beta_slice}), and their own Figures 1 and 3~\cite{brady2015thermoActivematter}, show these active agents are in phase equilibrium, since their effective utilities are equal. \\

This phase separation is not driven by thermodynamic quantities such as the chemical potential and Gibbs free energy, but rather by other factors that are relevant to the ``microscopic" mechanisms in the domain of the agents. These mechanisms could differ in different domains. They are different, for example, for bacteria in biology~\cite{venkat2022unified}, ants that ferry sand grains, mussels in the sea~\cite{liu2013CahnHilliard}, birds in a flock~\cite{venkat2022garuds}, and humans in social and economic settings~\cite{venkat2017book, venkat2024social}. But they all belong to the same \textit{universality class}, which is governed by the same mathematical structures and conditions described in Eq.~\ref{eq:utility-defn}-\ref{eq:ant-utility}, and Eq.~\ref{eq:schelling-utility} of statistical teleodynamics. This mathematical structure determines the properties of the arbitrage equilibrium. When the competition cost in the effective utilities of agents has the form $-\ln N_i$ (or $-\ln \rho)$, this arbitrage equilibrium corresponds to the thermodynamic or statistical equilibrium, as we discussed in Section V. \\

We see a similar correspondence for active Brownian particles whose velocities are density dependent, as discussed by Cates and Tailleur~\cite{cates2015motility}. They write the chemical potential $\mu$ of active particles without Brownian diffusion as
\begin{equation}
    \mu = \mu_{\mathrm{id}} + \mu_{\mathrm{ex}} = \ln \rho(\mathbf{r}) + \ln v(\rho (\mathbf{r}))
    \label{eq:mu-cates-velocity}
\end{equation}
where the propulsion speed $v(\rho)$ monotonically decreases with increasing density. In our framework, this would be equivalent to the effective utility 
\begin{eqnarray}
    h(\rho) = -\ln v(\rho (\mathbf{r})) - \ln\rho (\mathbf{r}) \label{eq:utility_v_dependence}
\end{eqnarray}

As Cates and Tailleur~\cite{cates2015motility} show, this dynamics will lead to phase separation if there exists a concave region in the free energy corresponding to Eq.~\ref{eq:mu-cates-velocity}. In our case, this would correspond to the existence of a convex region in the potential $\phi$ as seen in Eq.~\ref{eq:schelling-potential-equil} and Fig.~\ref{fig:SpinodalBinodal}. \\

However, a more complete version of this model should include the utility provided by the empty spaces in which the particles can potentially move. This becomes particularly important for high particle density $\rho$ when the empty space becomes valuable for motile particles. This is what we call an option benefit in Eq.~\ref{eq:schelling-utility}. Therefore, Eq.~\ref{eq:utility_v_dependence} now becomes
\begin{eqnarray}
    h(\rho) = -\ln v(\rho (\mathbf{r})) + \gamma \ln (1-\rho(\mathbf{r})) - \ln\rho (\mathbf{r}) 
    \label{eq:utility-chemo-option}
\end{eqnarray}

Furthermore, if the propulsion speed decays exponentially with respect to density, as Cates and Tailleur suggest~\cite{cates2015motility}, then we have $v(\rho) = \exp(-\alpha_v \rho)$, which leads to
\begin{eqnarray}
    h(\rho) = \alpha_v \rho + \gamma \ln (1-\rho(\mathbf{r})) - \ln\rho (\mathbf{r}) 
    \label{eq:utility-chemo-cubic}
\end{eqnarray}

For $\gamma = 1$, $\alpha_v > 4$ results in a non-monotonic cubic profile (i.e., the van der Waals loop), as seen in Fig.~\ref{fig:Utility_density_zero_beta}, and hence phase separation for the appropriate densities. 

\subsection{Connection with Chemotaxis and MIPS}

Zhao et al.~\cite{zhao2023chemotactic} discuss MIPS in the context of chemotaxis, where there is a directed motion of the active particle along the chemical gradient. While they predict and explain the effect of chemotaxis on MIPS from a dynamic perspective, here we do the same from the perspective of statistical teleodynamics. \\

Venkatasubramanian et al.~\cite{venkat2022unified} proposed the effective utility of a particle in a chemoattractant environment as 
\begin{eqnarray}
    h_i = \alpha c_i - \ln n_i
\end{eqnarray}
where $c_i$ is the concentration of the chemoattractant at a given state $i$, and $n_i$ is the number of particles at the state. The first term corresponds to an affinity for states (in this case, regions) with higher concentrations of the chemoattractant, and the second term corresponds to the entropic component as before. They also derive the game-theoretic potential, $\phi(\mathbf{x})$ as
\begin{eqnarray}
   \phi(\mathbf{x}) = \alpha \frac {C}{N} + \frac {1}{N} \ln \frac{N!}{\prod_{i=1}^n (Nx_i)!}
\end{eqnarray}

where $N$ is the number of active particles, $x_i = n_i/N$, and $C$ is the total amount of chemoattractant. This formulation is equivalent to that of O'Byrne and Tailleur~\cite{o2020lamellar}. \\

In the continuum limit, $c_i$ can be replaced by $c(\mathbf{r})$, the concentration distribution of the chemoattractant in the particle environment at the location $r$. Similarly, $N_i$ is replaced by $\rho(\mathbf{r})$, the density of the active particle at a given location. Now, the utility becomes  
\begin{eqnarray}
    h (c, \rho)= \alpha c(\mathbf{r}) - \ln \rho(\mathbf{r}) + \ln N \label{eq:continuum_chemotaxis}
\end{eqnarray}
and similarly, the game theoretic potential
\begin{eqnarray}
    \phi =  \frac{\int \alpha c(\mathbf{r}) \rho(\mathbf{r}) d\mathbf{r}}{N} - \frac{1}{N}\int \rho(\mathbf{r}) \ln \rho(\mathbf{r}) d \mathbf{r} \label{eq:potential-chemotaxis}
\end{eqnarray}

O'Byrne and Tailleur \cite{o2020lamellar}, who proposed a coarse-grained diffusive model for active matter dynamics driven by a \textit{chemorepellant} concentration field $c$, describe the resultant coarse-grain dynamics using an effective free energy functional $\mathcal{F}$ and deterministic flux $J_D$.
\begin{eqnarray}
J_D = -D \rho \nabla \left\{\left[v_1 + v_0\frac{\alpha_1 + (d-1)\Gamma_1}{\alpha_0 + (d-1)\Gamma_0}\right]\frac{c}{d D} + \log \rho \right\} \nonumber\\\label{eq:jd-o2020}
\end{eqnarray}
where $J_D = \nabla \left(\delta\mathcal{F}/\delta \rho\right)$, the gradient of the functional derivative of the free energy ($\rho$ is the particle density).  

\begin{eqnarray}
\delta\mathcal{F}/\delta \rho = -D \left\{\left[v_1 + v_0\frac{\alpha_1 + (d-1)\Gamma_1}{\alpha_0 + (d-1)\Gamma_0}\right]\frac{c}{d D} + \log \rho \right\} \nonumber\\\label{eq:Fder-o2020}
\end{eqnarray}

We identify that this functional derivative, $\left(\delta\mathcal{F}/\delta \rho\right)$, is the chemical potential of the active particle, which is the negative of the utility in Eq.~\ref{eq:continuum_chemotaxis}. Once again, as before, we see the equivalence of these two frameworks. Observe that our  Eq.~\ref{eq:continuum_chemotaxis} maps to Eq.~\ref{eq:Fder-o2020} upto a proportionality constant with  $\alpha= -\left[v_1 + v_0\dfrac{\alpha_1 + (d-1)\Gamma_1}{\alpha_0 + (d-1)\Gamma_0}\right]\dfrac{1}{dD}$. The simulations performed in the study (with the parameter values: $v_0=1, v_1 = 0.2, \alpha_0 = 50, \alpha_1 = \Gamma_1=0$) correspond to $\alpha<0$ in our formulation, the chemorepellant case. \\

If the initial amount of the chemoattractant is fixed, as the particles consume the chemoattractant, its local concentration decreases with increasing local density $\rho(r)$ of the particles. This decrease can be modeled as $c(\mathbf{r}) = -k' \rho[\mathbf{r}]$, giving density-dependent utility as,
\begin{eqnarray}
    h (\rho)= -\alpha' \rho- \ln \rho + \text{constant}
    \label{eq:utility-chemotaxis-rho}
\end{eqnarray}

where $\alpha' = \alpha k'$. Now, consider the density-dependent velocity model of the active Brownian particles. Incorporating that dynamics (Eq.~\ref{eq:utility-chemo-cubic}) into Eq.~\ref{eq:utility-chemotaxis-rho}, we have 
\begin{eqnarray}
    h (\rho)&=& (\alpha_v - \alpha') \rho(\mathbf{r}) + \gamma\ln (1-\rho(\mathbf{r})) - \ln\rho(\mathbf{r})
    \label{eq:utility-alpha-alpha'}
\end{eqnarray}

Rewriting Eq.~\ref{eq:utility-alpha-alpha'} in terms of the chemical potential ($\mu^{act}$), we have 
\begin{eqnarray}
    \mu^{act} (\rho)&=& - (\alpha_v - \alpha') \rho(\mathbf{r}) - \gamma\ln (1-\rho(\mathbf{r})) + \ln\rho(\mathbf{r}) \nonumber\\
    \label{eq:mu-alpha-alpha'}
\end{eqnarray}

\begin{figure}[!ht]
    \centering
    \includegraphics[width=0.5\textwidth]{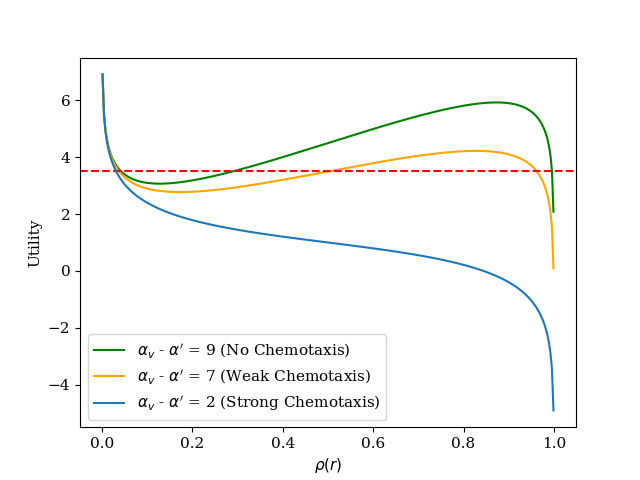}
        \caption{{\textbf{Effect of chemotaxis on utility based on Eq. \ref{eq:utility-alpha-alpha'} with $\gamma=1$}}}
    \label{fig:chemotaxis-h-three}
\end{figure}

This system can show phase separation depending on the $\alpha_v$ and $\alpha'$ values. We show, for example, three cases ($\gamma =1)$: (i) no chemotaxis: $\alpha_v = 9$ and $\alpha' = 0$, (ii) weak chemotaxis: $\alpha_v = 9$ and $\alpha' = 2$, and (iii)  strong chemotaxis: $\alpha_v = 9$ and $\alpha' = 7$ in Fig.~\ref{fig:chemotaxis-h-three}. We see that the non-monotonocity of $h$ changes depending on the values of $\alpha_v$ and $\alpha'$, thus determining whether a phase separation occurs or not. Specifically, the nonmonotonic cubic profile (i.e., the van der Waals loop) occurs when $\alpha_v - \alpha' > 4$. We see that a strong presence of chemotaxis can prevent phase separation (blue curve). 

\section{Conclusions}

In developing a theory of emergent behavior of self-propelled agents that dynamically change states, one faces three questions from the very beginning: (i) Why do the agents change states? (ii) How do they change states? (iii) What happens to the system, i.e., the population of such agents, eventually? \\

The answers to the first two questions depend on the ``microscopic" fundamental mechanisms of the particular situation. For example, gas molecules change states driven by thermal agitation, ants by pheromone signals, flocking birds by visual and auditory cues, humans by socioeconomic considerations, and so on. Thus, the ``microscopic" details of why and how agents move around depend on the appropriate fundamental mechanisms of physics, chemistry, biology, ecology, sociology, or economics, as the case may be. Statistical teleodynamics is agnostic of such ``microscopic" mechanisms. It has a ``mesoscopic" view of the agents, as it answers the question of \textit{what} happens eventually. \\

Towards this goal, we have developed simple models that offer an appropriate coarse-grained description that is not restricted by system-specific details and nuances, but without losing key conceptual insights and relevance to empirical macroscopic phenomena. The spirit of our modeling is similar to that of van der Waals in developing his equation of state. \\

In our theory, the central concept is the effective utility $h_i$ of an agent. All agents constantly compete to increase their utilities in an environment with resource constraints. The resources may be energy, space, food, money, etc. This competition, under certain conditions, eventually leads to an arbitrage equilibrium in which all agents have the same effective utility. \\

Given the ``mesoscopic" coarse-graining in our models, the effective utility $h_i$ is agnostic of the ``microscopic" details of its components. Consider, for example, the emergent behaviors of ants and Janus particles. Their ``microscopic" mechanisms of motion, i.e., the ``whys" and ``hows," are very different for the two cases. However, these different ``microscopic" mechanisms result in the same ``mesoscopic" model of their effective utilities (Eq.~\ref{eq:ant-utility} and Eq.~\ref{eq:janus-utility}), thereby predicting the same ``macroscopic" emergent behavior, namely, the Weibull distribution. \\

We find a similar situation with motility-induced phase separation phenomena. As our mathematical analysis shows in Section V, the necessary and sufficient conditions for phase separation are, respectively: (i) the effective utility function at arbitrage equilibrium ($h^*$) must be at least a cubic or cubic-equivalent of the order parameter (e.g., $\rho$ in Eq.~\ref{eq:schelling-utility_equil}) with three distinct real and positive zeros, and (ii) the initial value of the order parameter is within the spinodal region of the miscibility gap (e.g., Figs.~\ref{fig:SpinodalBinodal}-\ref{fig:phase_separation}).  \\

Again, these ``mesoscopic" requirements can be met by using many different ``microscopic" mechanisms. Janus particles in the work of Takatori et al.~\cite{brady2015thermoActivematter} do it one way, but mussels in the sea use different mechanisms for their pattern formation~\cite{liu2012alternativemechanisms}. In sociology and economics, agents use entirely different ``microscopic" socioeconomic processes to dynamically switch states that lead to social and economic segregation~\cite{venkat2017book, schelling1969segregation, venkat2024social}. \\

Comparing our model in Eq.~\ref{eq:schelling-utility_equil} and Eq.~\ref{eq:schelling-potential-equil} with that of Takatori et al.~\cite{brady2015thermoActivematter} in Eq.~\ref{eq:pi-act}-\ref{eq:mu-act}, we see that the latter is more complicated, as it reflects the ``microscopic" details of this dynamics. In our simpler model, such details have been avoided. However, their phase diagram (Fig. 1 in \cite{brady2015thermoActivematter}) and the Gibbs free energy vs. volume fraction plot
(Fig. 3 in \cite{brady2015thermoActivematter}) have the same qualitative features as our model shown in Fig.~\ref{fig:SpinodalBinodal}-\ref{fig:phase_separation}. These qualitative features are the existence of two minima in the Gibbs free energy (corresponding to two maxima in our game-theoretic potential $\phi$) and the miscibility gap in the phase diagram. \\

Although these motility-induced phase separation phenomena are generally characterized as nonequilibrium or out-of-equilibrium phenomena~\cite{marchetti2013hydrodynamics, cates2015motility, brady2015thermoActivematter}, our analysis recognizes them as the result of the arbitrage equilibrium. As we discussed in Section V, this is, in principle, the same as thermodynamic equilibrium, with the only difference being the arbitrage currency. In thermodynamics, the currency is the chemical potential, and in teleodynamics, it is the effective utility. Otherwise, the mathematical structure in both situations is the same. This resolves the puzzle noted by many~\cite{berkowitz2020active,o2020lamellar, cates2013active, cates2015motility, gonnella2015motility, brady2015thermoActivematter} that some active matter systems that look like out-of-equilibrium systems at the microscopic scale behave macroscopically like simple equilibrium systems of passive matter. \\

Statistical teleodynamics is applicable to the entire range of the agency spectrum going from purpose-free thermodynamic agents (e.g., molecules) to purpose-driven rational agents (e.g., humans) with the different kinds of biological and ecological agents (e.g., bacteria, ants, birds, mussels, etc.) situated somewhere in between in this spectrum of self-actualizing capabilities. This is summarized in Table \ref{tab:utility-domains}, which lists utility function templates in different domains. In this paper, we have already discussed examples of the thermodynamic game, ant-crater formation, and social segregation. In economics, the emergence of the income distribution can be modeled by~\cite{venkat2015howmuch, venkat2017book}
\begin{equation}
h_{i}= \alpha \ln S_i - \beta \left(\ln S_i\right)^2 - \ln N_i.
\end{equation}

where the first term is the benefit of income ($S_i$), the second is the cost of work expended to earn this income, and the last is the cost of competition.\\

In ecology, the flocking behavior of bird-like agents is modeled by~\cite{venkat2022garuds}
\begin{equation}
h_{i}= \alpha N_i - \beta N_i^2+ \gamma N_i l_i - \ln N_i
\end{equation}

where the first term is the security benefit of having many birds in the neighborhood, the second is the congestion cost of these neighbors, the third is the alignment ($l_i$) benefit of flying in the same direction as the neighbors, and the last is again the cost of competition. \\

We wish to stress that all of these systems reach arbitrage equilibria. Some of the equilibria have well-known distributions as outcomes, such as exponential (energy), lognormal (income), or Weibull (ant craters, Janus particles). But some others have ``messy" distributions, as in the case of social segregation and birds flocking. The specific outcome depends on the non-entropic terms in the effective utility function, i.e., the ``first law" of teleodynamics terms that enforce the constraints on entropy maximization. \\

\begin{table}[!h]
    \centering
    \caption{Utility functions in different domains}
    \label{tab:utility-domains}
    \scalebox{0.9}{\begin{tabular}{c c c }\\
    \hline
        \textbf{Domain}& \textbf{System}& \textbf{Utility function} ($h_i$) \\ \hline\\
         Physics & Energy distribution & $-\beta  E_i - \ln N_i$ \\\\
         Physics & Janus particles & $- \dfrac{\omega r_i^a}{a} - \ln N_i$ \\\\
         Biology & Bacterial chemotaxis & $\alpha c_i - \ln N_i$ \\\\
         Ecology & Ant craters & $b - \dfrac{\omega r_i^a}{a} - \ln N_i$ \\\\
         Ecology & Birds flocking & $\alpha N_i - \beta N_i^2+ \gamma N_i l_i - \ln N_i$ \\\\
         Sociology & Social segregation  & $\eta N_i - \xi N_i^2 + \ln(H-N_i) -\ln N_i$ \\\\
         Economics & Income distribution & $\alpha \ln S_i - \beta \left(\ln S_i\right)^2 - \ln N_i$ \\\\
        \hline
    \end{tabular}}
\end{table}

By comparing the mathematical structure of the various effective utility functions, we observe a certain \textit{universality} across the different domains. They are all based on benefit-cost trade-offs, but the actual nature of the benefits and costs depend on the details of the specific domain, as one would expect. The main requirement to belong to this universality class is that the disutility due to competition can be modeled (or at least reasonably approximated) as $-\ln N_i$ (discrete case) or $-\ln \rho$ (continuous case). This is a critical requirement, as Kanbur and Venkatasubramanian explain~\cite{kanbur2020occupational}. This agent-based property directly leads to the system-wide property of entropy, thereby connecting the agents and the system in a cohesive mathematical framework. This term also facilitates the integration of potential game theory with statistical mechanics, paving the way for a universal theory of emergent equilibrium phenomena in active and passive matter. \\

One might argue that our theory is not that different from what has already been proposed to explain MIPS using chemical the potential and free-energy-based approaches reported in the literature~\cite{cates2015motility, brady2015thermoActivematter}. We agree that the conventional thermodynamic perspective is suitable for physicochemical systems, although even here, one has to invoke the new concept of ``nonequilibrium chemical potential"~\cite{cates2015motility, brady2016MipsCurrent}. Thus, conventional concepts based on thermodynamics are already proving inadequate for handling even physicochemical systems. We believe that it would be even more problematic for higher-order living agents, such as those listed in Table \ref{tab:utility-domains}.  For example, how would one relate the salary $S_i$ of an economic agent to its ``nonequilibrium chemical potential"? A utility-based perspective is much more intuitive and, hence, a natural framework. The other important contribution, we believe, is the conceptual progress we have made in recognizing that many of these so-called nonequilibrium systems are actually systems in equilibrium, an \textit{arbitrage} equilibrium. \\

Our analysis suggests that the pursuit of maximum utility or survival fitness, combined with competitive dynamics under constraints, could be a universal self-organizing mechanism for active matter. In biology, in general, the search for improving one's fitness occurs in the design space of genetic features. Here, mutation and crossover operations facilitate movements in the feature space such that an agent improves itself genetically via Darwinian evolution to increase its utility, i.e., survival fitness. In economics, on the other hand, agents search in the products and/or services space so that they can offer better products/services to improve their economic survival fitness in a competitive marketplace. \\

Thus, this mechanism is reminiscent of Adam Smith's \textit{ invisible hand}. In all these different domains, each agent pursues its own self-interest to increase its own $h_i$, but a stable collective order emerges spontaneously as a result of the competitive dynamics among all agents under constraints. Therefore, we suggest that it is a \textit{pair} of invisible hands: one is the pursuit of maximum utility and the other the competitive dynamics under constraints. We need both principles for the arbitrage equilibrium to emerge spontaneously. \\

By formulating equilibria in active matter more broadly, unrestricted by the narrow confines of thermodynamics, statistical teleodynamics and arbitrage equilibria open up conceptual possibilities that coherently accommodate active matter in the entire range of nonliving to living agents more naturally. However, what we have presented is a van der Waals-like version of statistical teleodynamics. Much more needs to be done to further develop this framework to address challenging emergent phenomena in physics, chemistry, biology, ecology, sociology, and economics. 

\section*{Methods}
The agent-based simulation was performed using Python. We distributed agents on a 2-D $ 300 \times 300$ grid with 90,000 cells. Three simulation studies are reported in this paper -- with 22,500 agents, 45,000 agents, and 55,000 agents. For each case, initially, the agents were randomly distributed on the grid, with each agent occupying one cell. \\

The dynamical evolution of the system is determined by two neighborhoods around an agent $i$. One is the local neighborhood of \emph{interaction}, which is an area with 49 cells that surround the agent $i$ (including the cell $i$ is occupying). The other is the \emph{exploration} neighborhood (which is larger than the interaction neighborhood and contains it) within which an agent $i$ can explore and move to another cell to improve its utility $h_i$. The exploration neighborhood has 1680 cells. The neighborhood sizes are parameters that can be varied to balance the need to allow for complex patterns to emerge at arbitrage equilibrium and the need to accomplish this in a reasonable amount of computational time. We found that our combination (49 and 1680) accomplishes this well. \\

The density of agents in any cell is defined as the ratio of the number of agents in the interaction neighborhood to the total number of cells in the neighborhood. At each iteration, every agent is given the opportunity to move to a vacant cell in the exploration neighborhood where it would have higher utility than its current cell. If the agent does not find a vacant cell, it chooses to stay at its current location. After an agent moves, its utility, and its neighbors' density and utility are updated. The simulations were carried out for 10,000 iterations, at which time the system typically reached the arbitrage equilibrium. 

\section*{Acknowledgements}
The first author thanks John Brady, Kyle Bishop, and Chris Durning for helpful discussions. This work was supported in part by a grant to the Center for Managing Systemic Risk from Columbia University.

\section*{Author Contributions}
VV: Conceptualization, Theory, Methodology, Analysis, Investigation, Supervision, Funding Acquisition, and Writing; A. Sivaram: Methodology, Analysis, Investigation, and Writing; NS: Software development and analysis; A. Sankar: Software development and analysis.

The authors have no conflicts of interest to declare.

\section*{Data Availablity and Reproducibility Statement}
Data for Fig. 1-4 are from Takatori et al. [18]. Figures 5-6 are plotted using Eq. 19. Figures 7-10 are plotted using Eq. 22. Figures 11-12 are from agent-based simulations, the code of which will be made available upon request to the corresponding author. We are currently updating the code for better ease of use. Figure 13 is plotted using the equation. 32-33, and Figure 14 using Eq. 48.

%

\end{document}